\documentclass{llncs}
\usepackage{graphicx}
\usepackage{bigdelim}
\usepackage{multirow}
\usepackage{amssymb,amsmath}
\usepackage{amsfonts}
\usepackage{framed}
\usepackage{pifont}
\usepackage{wrapfig}
\usepackage{color}
\usepackage{pdfpages}
\usepackage{listings}
\usepackage{enumitem}
\usepackage{url}
\usepackage{breakurl}
\usepackage{caption}
\usepackage{graphicx}
\usepackage{subcaption}

\setitemize{partopsep=0pt, parsep=0pt, itemsep=0pt}

\usepackage[utf8]{inputenc}

\RequirePackage[bookmarks,
                unicode,
                urlcolor=black,
                citecolor=black,
                linkcolor=black,
                pagecolor=black,
                colorlinks,
                hyperfigures]{hyperref}

\renewcommand{\tablename}{~\vspace{5mm}\\Tab.}

\usepackage{fancyhdr}
\fancypagestyle{firststyle}
{
  \fancyhf{}
  \fancyfoot[L]{\scriptsize{This work is a result of the lab course ``Business Intelligence II (2018W)'' and has been partially supported and funded by the Austrian Research Promotion Agency (FFG) via the ``Austrian Competence Center for Digital Production'' (CDP) under the contract number 854187.}}
}

\title{Combining Conformance Checking and Classification of XES Log Data For The Manufacturing Domain}
\author{Matthias Ehrendorfer, Juergen-Albrecht Fassmann, Juergen Mangler, Stefanie Rinderle-Ma}
\institute{University of Vienna, Vienna, Austria,\\matthias.ehrendorfer@gmail.com, juergen.fassmann@gmail.com\\~\\University of Vienna, Faculty of Computer Science, \\Research Group Workflow Systems and Technology, Vienna, Austria,\\\{juergen.mangler, stefanie.rinderle-ma\}@univie.ac.at}

\begin{document}
{\def\addcontentsline#1#2#3{}\maketitle}
\thispagestyle{firststyle}

\section{Introduction} 
\label{sec:intro}

Currently, data collection on the shop floor is based on individual resources
such as machines, robots, and Autonomous Guided Vehicles (AGVs). There is a gap
between this approach and manufacturing orchestration software that supervises
the process of creating single products and controls the ressources'
interactions. This creates the need to save resource-based data streams in
databases, clean it, and then re-contextualize it, i.e., by connecting it to
orders, batches, and single products. Looking at this data from a
process-oriented analysis point of view enables new analysis prospects. This
paper utilises these prospects in an experimental way by creating BPMN models
for the manufacturing of two real-world products: (1) a low volume, high
complexity lower-housing for a gas-turbine and (2) a high volume, low
complexity, small tolerance valve lifter for a gas turbine. In contrast to the
resource-based data collection, 30+ values are modeled into the BPMN models and
enacted by a workflow engine, creating execution logs in the XES standard
format. Conformance checks are carried out and interpreted for both scenarios
and it is shown how existing classification and clustering techniques can be
applied on the collected data in order to predict good and bad parts, ex-post
and potentially at run-time.

\begin{figure}[h]
  \centering
  \includegraphics[width=0.5\textwidth]{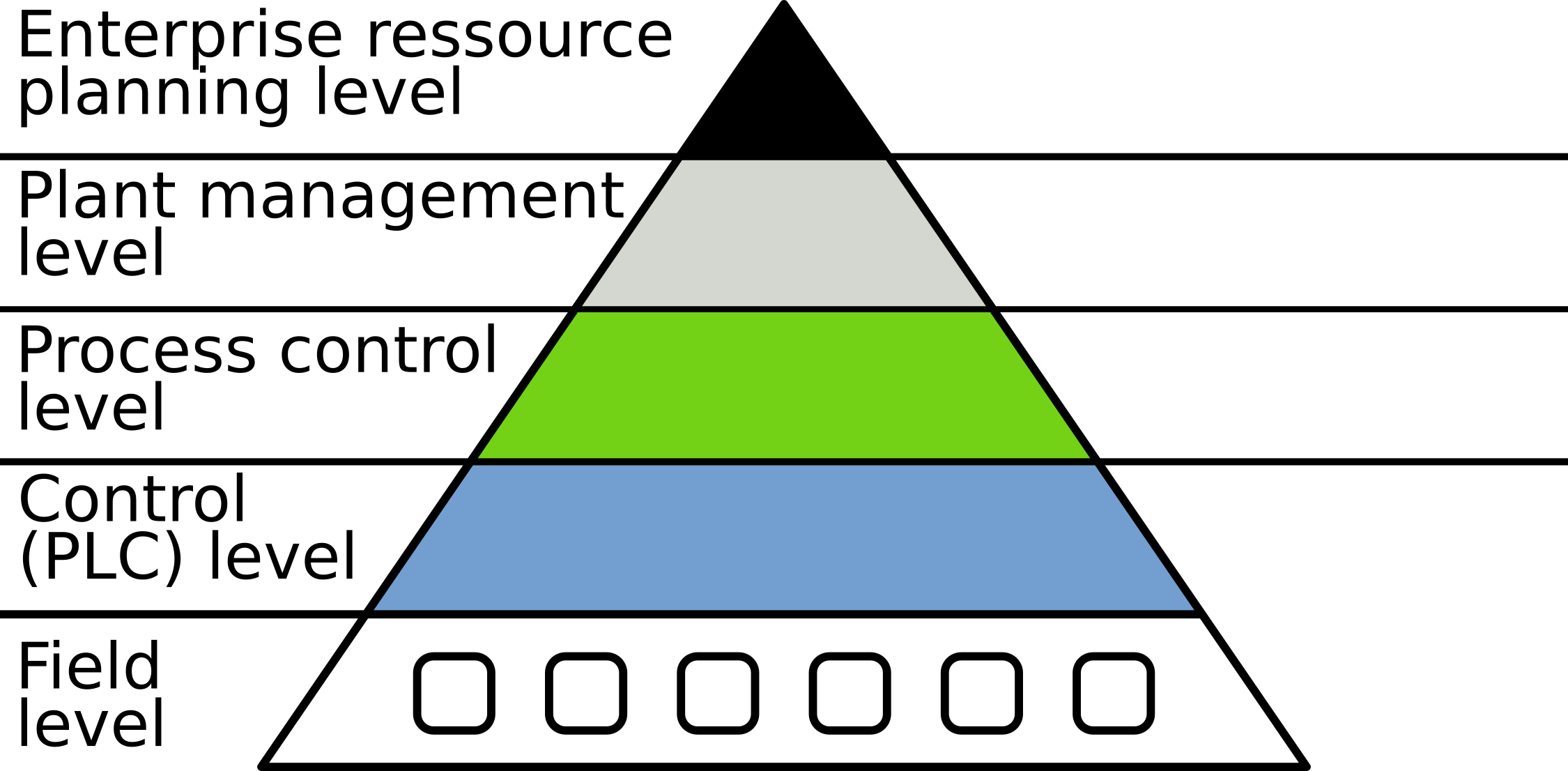}
  \caption{Automation Pyramid}
  \label{fig:automationpyramid}
\end{figure}

The process created for the manufacturing of both parts can be divided into a
number of subprocesses which correspond to different levels of the automation
pyramid (given in Fig. \ref{fig:automationpyramid}). This separation allows a
better overview as well as special foci in each of the subprocesses. The
subprocesses are:

\begin{itemize}

  \item The ``Order Processing'' process which is responsible for detecting the
  start of the production and spawning a subprocess for every part produced.
  Therefore, this process connects a number of part production processes for an
  order.

  \item The ``Part Production'' process describes the process steps as well as
  the order in which they are executed on the level of production or measuring
  steps and spawns subprocesses for each of the production steps.

  \item The ``Production'' process handles the detection of individual machines
  starting and finishing machining and spawns subprocesses for data collection
  while waiting for the production step to finish.

  \item The ``Machining'' process is responsible for capturing data sent by
  individual machines during machining.

\end{itemize}

The ``Order Processing'' process corresponds to the ``Plant management level''
of the automation pyramid while the ``Production Process'' resides on the
``Process control level'' and the ``Machining Process'' can be assigned to the
``Control (PLC) level''. The ``Part Production'' process is the link between
``Order Processing'' and ``Production'' and can therefore be found between
level 2 and 3 (i.e. the second and third one counting from the top) of the
automation pyramid.

When executing the processes, log files are created. They are available as YAML
files in the XES standard format and contain the tasks given in the BPMN models
which are described in XML format. Production data obtained during the
machining as well as measurement results are also included in the corresponding
task of the log files. Due to the abovementioned usage of subprocesses, a
number of log files have to be merged in order to obtain all the process steps
needed for the production of one part. The information to do this (i.e. which
subprocess is spawned) is also contained in the task creating the subprocess of
the parent process.


As described above, the part produced in the first scenario is a low volume,
high complexity lower-housing for a gas-turbine. Therefore, the parts produced
are measured during the production process and repaired if they do not comply
to the specifications of the plan. This leads to a higher number of machining
processes as machining might be interrupted and continued a number of times.
The log data for the first scenario can be download from
\footnote{\url{http://cpee.org/~demo/bus_paper/data/lowerhousing.tgz}}.


The part produced in the second scenario is a high volume, low complexity,
small tolerance valve lifter for a gas turbine. In contrast to the part
produced in the first scenario it is not reasonable to repair bad parts and
therefore, after the machining process manual as well as automatic measurement
is performed but no repair is carried out. Due to an error while performing the
manual measurement, the manual measurement results of parts 129-180 are shifted
by 2. The log data for the second scenario can be downloaded from
\footnote{\url{http://cpee.org/~demo/bus_paper/data/gv12.tgz}}.


\section{Conformance Checking of Shopfloor Processes} 
\label{sec:conf}

\subsection{Modelling the Processes}
\label{sec:confmodelling}
In order to perform conformance checking, the logs as well as the templates need to be transformed to a format that allows easy checking. The ProM tool (for further information see \cite{prom}) provides the ``PNetReplayer'' (\cite{pnetreplayer}) package which allows performance checking using the ``Replay a Log on Petri Net for Conformance Analysis'' function. This function takes a petri net (in a TPN format) and a XES log file which is based on XML as input files and creates a result file containing fitness values and the replay results for each log. Therefore, the BPMN process template files given as XMLs (for further information see section \ref{sec:intro}) need to be transformed to petri nets. The transformation from the BPMN process template to the petri net is performed using the following rules:

\begin{itemize}
  \item Each ``call'' element is transformed into two transitions with a place in between and afterwards. The first transition is identifiable as start and the second one as end which is later used to link them to the start or complete lifecycle transitions in the logs represented as XES file.
  \item Each ``manipulate'' element is transformed into a transition with a place afterwards. As the log contains only one event for such elements there is no need to create a start and end transition.
  \item ``terminate'' elements are transformed to a transition with a place that has no outgoing edges afterwards.
  \item ``loop'' elements are transformed to two transitions (starting and closing) with a place after each of them and the second place being connected to the first transition in addition to the transition created by the element after the loop. As a loop contains other elements, the content between these two transitions is defined by the contained elements.
  \item ``choose'' elements are transformed into two places where the first one splits into a number of branches and the second one merges these branches. ``alternative'' and ``otherwise'' are both handled the same way as it does not make a difference if one of the branches is the default one when using petri nets for conformance checking. Therefore, an ``alternative'' or ``otherwise'' element is transformed into two transitions where the first one has an incoming edge from the first place created by the ``choose'' element and is followed by a place and the second one is connected to the merging place created by the ``choose'' element described before. As a branch contains other elements, the content of a branch is defined by the contained elements.
  \item ``parallel'' elements are transformed to two transitions. The first one is connected to a number of places equal to the number of parallel branches via edges. The second transition merges these parallel branches by having incoming edges from the last place of each parallel branch. The ``parallel\_branch'' element is transformed to a transition signalling the start of the parallel branch with a place afterwards. As a branch contains other elements, the content of a branch is defined by the contained elements.
\end{itemize}

As a prerequisite for the transformation with the code provided, the process templates need to be in the XML format and contain only the elements defined in \cite{mangler2010cloud}. Additionally, the XML element ``/testset/description/description'' must be available and contain the elements mentioned above with the defined attributes and a valid nesting structure. Using the rules given before, the BPMN processes are then transformed to petri nets using the TPN file format.

The program implementing the rules was written in python (the code is also available at \cite{scenario2_processToTpn} for scenario 2 - at \cite{scenario1_processToTpn} for scenario 1 - and can be executed by using the following command: ``./processToTpn.py [path to existing XML] [path to TPN that should be created]''):

\lstset{language=Python}
\begin{lstlisting}
#!/usr/bin/python3

import xml.etree.ElementTree as ET;
import sys;
import re;

counter=0;
termination=False;
startstack=[];
endstack=[];
loopstack=[];
printString='';

def processChild(node, indent, firstOfBranch):
  global counter;
  global startstack;
  global endstack;
  global loopstack;
  global printString;
  global termination;
  exception=False;
  indentString='';
  i=0;
  while i<indent:
    indentString+=' ';
    i+=1;
  if 'parallel_branch' in node.tag:
    print(indentString+node.tag);
    print('trans '+'x_parallel_branch');
    printString+=('trans '+'x_parallel_branch');
    printString+='\n';
    x=startstack.pop();
    print('  in '+x);
    printString+=('  in '+x);
    printString+='\n';
    counter+=1;
    print('  out p'+str(counter)+';');
    printString+=('  out p'+str(counter)+';');
    printString+='\n';
    #hinzufuegen pfad zwischen transition von parallel und (neuem) place
  elif 'parallel' in node.tag:
    print(indentString+node.tag);
    print('trans '+'x_parallel');
    printString+=('trans '+'x_parallel');
    printString+='\n';
    print('  in p'+str(counter));
    printString+=('  in p'+str(counter));
    printString+='\n';
    i=0;
    outString='  out';
    while i<len(list(node)):
      counter+=1;
      outString+=' p'+str(counter);
      startstack.append('p'+str(counter));
      if (i+1)<len(list(node)):
        outString+=',';
      i+=1;
    outString+=';';
    print(outString);
    printString+=(outString);
    printString+='\n';
    #hinzufuegen einer transition
  elif 'loop' in node.tag:
    print(indentString+node.tag);
    loopstack.append('p'+str(counter));
    #hinzufuegen eines pfades zwischen dieser transition und dieser transition
  elif 'choose' in node.tag:
    print(indentString+node.tag);
    i=0;
    while i<len(list(node)):
      startstack.append('p'+str(counter));
      i+=1;
    #counter+=1;
    #hinzufuegen eines places
  elif 'alternative' in node.tag:
    print(indentString+node.tag);
    print('trans '+'x_alternative');
    printString+=('trans '+'x_alternative');
    printString+='\n';
    x=startstack.pop();
    print('  in '+x);
    printString+=('  in '+x);
    printString+='\n';
    counter+=1;
    print('  out p'+str(counter)+';');
    printString+=('  out p'+str(counter)+';');
    printString+='\n';
    endstack.append('p'+str(counter));
    exception=True;
    print(endstack);
    #hinzufuegen pfad zwischen place von choose und (neuer) transition
  elif 'otherwise' in node.tag:
    print(indentString+node.tag);
    print('trans '+'x_alternative');
    printString+=('trans '+'x_alternative');
    printString+='\n';
    x=startstack.pop();
    print('  in '+x);
    printString+=('  in '+x);
    printString+='\n';
    counter+=1;
    print('  out p'+str(counter)+';');
    printString+=('  out p'+str(counter)+';');
    printString+='\n';
    endstack.append('p'+str(counter));
    exception=True;
    print(endstack);
    #hinzufuegen pfad zwischen place von choose und (neuer) transition
  elif 'call' in node.tag:
    print(indentString+node.tag);
    print('trans '+node[0][0].text.replace(' ','').replace('?','')+'_'+node.attrib['id']+'_'+endpointToUrl[node.attrib['endpoint']]+'_start');
    printString+=('trans "'+node[0][0].text.replace(' ','').replace('?','')+'_'+node.attrib['id']+'_'+endpointToUrl[node.attrib['endpoint']]+'_start"');
    printString+='\n';
    print('  in p'+str(counter));
    printString+=('  in p'+str(counter));
    printString+='\n';
    counter+=1;
    print('  out p'+str(counter)+';');
    printString+=('  out p'+str(counter)+';');
    printString+='\n';

    print('trans '+node[0][0].text.replace(' ','').replace('?','')+'_'+node.attrib['id']+'_'+endpointToUrl[node.attrib['endpoint']]+'_complete');
    printString+=('trans "'+node[0][0].text.replace(' ','').replace('?','')+'_'+node.attrib['id']+'_'+endpointToUrl[node.attrib['endpoint']]+'_complete"');
    printString+='\n';
    print('  in p'+str(counter));
    printString+=('  in p'+str(counter));
    printString+='\n';
    counter+=1;
    print('  out p'+str(counter)+';');
    printString+=('  out p'+str(counter)+';');
    printString+='\n';

    if not firstOfBranch:
      endstack.pop();
    endstack.append('p'+str(counter));
    print(endstack);
    #counter+=1;
    #hinzufuegen von transition mit id
  elif 'manipulate' in node.tag:
    print(indentString+node.tag);
    print('trans '+node.attrib['label'].replace(' ','').replace('?','')+'_'+node.attrib['id']);
    printString+=('trans '+node.attrib['label'].replace(' ','').replace('?','')+'_'+node.attrib['id']);
    printString+='\n';
    print('  in p'+str(counter));
    printString+=('  in p'+str(counter));
    printString+='\n';
    counter+=1;
    print('  out p'+str(counter)+';');
    printString+=('  out p'+str(counter)+';');
    printString+='\n';
    if not firstOfBranch:
      endstack.pop();
    endstack.append('p'+str(counter));
    print(endstack);
    #counter+=1;
    #hinzufuegen von transition mit id
  elif 'terminate' in node.tag:
    print(indentString+node.tag);
    print('trans '+'x_termination');
    printString+=('trans '+'x_termination');
    printString+='\n';
    print('  in p'+str(counter));
    printString+=('  in p'+str(counter));
    printString+='\n';
    counter+=1;
    print('  out p'+str(counter)+';');
    printString+=('  out p'+str(counter)+';');
    printString+='\n';
    if not firstOfBranch:
      endstack.pop();
    endstack.append('p'+str(counter));
    termination=True;



  if len(list(node))>0:
    x=1;
    for child in node:
      if (x == 1) and (not exception):
        processChild(child, indent+2, True);
      else:
        processChild(child, indent+2, False);
      x+=1;
  else:
    #print('no childs');
    pass;

  if 'parallel_branch' in node.tag:
    pass;
  elif 'parallel' in node.tag:
    counter+=1;
    print('trans '+'x_closing_parallel');
    printString+=('trans '+'x_closing_parallel');
    printString+='\n';
    i=0;
    inString='  in';
    while i<len(list(node)):
      inString+=' '+endstack.pop();
      if (i+1)<len(list(node)):
        inString+=',';
      i+=1;
    outString+=';';
    print(inString);
    printString+=(inString);
    printString+='\n';
    print('  out p'+str(counter)+';');
    printString+=('  out p'+str(counter)+';');
    printString+='\n';
    endstack.append('p'+str(counter));
    #close parallel
  elif 'choose' in node.tag:
    counter+=1;
    i=0;
    if termination:
      i+=1;
      termination=False;
    while i<len(list(node)):
      print('trans '+'x_closing_decision');
      printString+=('trans '+'x_closing_decision');
      printString+='\n';
      x=endstack.pop();
      print('  in '+x);
      printString+=('  in '+x);
      printString+='\n';
      print('  out p'+str(counter)+';');
      printString+=('  out p'+str(counter)+';');
      printString+='\n';
      i+=1;
    endstack.append('p'+str(counter));
    #close decision
  elif 'loop' in node.tag:
    print('trans '+'x_closing_loop');
    printString+=('trans '+'x_closing_loop');
    printString+='\n';
    print('  in p'+str(counter));
    printString+=('  in p'+str(counter));
    printString+='\n';
    x=loopstack.pop();
    print('  out '+x+';');
    printString+=('  out '+x+';');
    printString+='\n';
    #close loop
      
      
print('hello world');

endpointToUrl={};
tree = ET.parse(sys.argv[1]);
root = tree.getroot();

for topchild in root:
    x=1;
    if(topchild.tag == 'description'):
      for child in topchild[0]:
        if x == 1:
          processChild(child, 0, True);
        else:
          processChild(child, 0, False);
        x+=1;
    elif(topchild.tag == 'endpoints'):
      for child in topchild:
          regex=re.match('(\{.*\})(.*)', child.tag);
          print(regex.group(2)+'  '+(child.tag)+'  '+str(child.text));
          endpointToUrl.update({regex.group(2):child.text});

print(startstack);
print(endstack);
print(loopstack);
#print(printString);

completePrintString='';
i=0;
while i<counter:
  completePrintString+='place p'+str(i);
  if i==0:
    completePrintString+=' init 1'
  completePrintString+=';\n';
  i+=1;

completePrintString+=printString;
print(completePrintString);

with open(sys.argv[2], 'w') as file:
  print(completePrintString, file=file);

print(endpointToUrl);

\end{lstlisting}

In order to create the XES log file, all log files given in YAML format as described in section \ref{sec:intro} have to be transformed. To achieve this, the information from the YAML ``event'' elements containing a ``cpee:lifecycle:transition'' element with the value ``activity/calling'' or ``activity/done'' is written into the XES file. Depending on the ``cpee:lifecycle:transition'' the events are considered to be either start or end of the task. The final XES file contains all events (starting as well as ending) of all logs.

The XES file is structured into a header section, which includes extensions, global keys for events and traces and classifiers, which are used to classify the events. An additional classifier is added for the second scenario, which is the endpoint in combination with the CPEE lifecycle transition. The reason for this is explained in section \ref{sec:confscenario2}. Following, are multiple traces, one for each YAML log and each trace contains multiple events, which are, as just mentioned, mapped to the starting and completing of tasks in the YAML log file. Each trace has an ID or ``concept:name'', a name or ``cpee:name'' and the UUID. Each event has a name, an endpoint (although not all events actually have an endpoint, namely all ``script'' tasks), an ID (which is the ID of the task in the template), the lifecycle transition (``start'' or ``complete''), the CPEE lifecycle transition (``activity/calling'' and ``activity/done'') and the timestamp.

The program transforming the YAML logs to a XES file as described above is written in python (see also \cite{scenario2_xes_map} for scenario 2 and \cite{scenario1_xes_map} for scenario 1) and can be executed by using the following command: ``python xes\_map.py'' - in order to execute successfully the slightly edited library XES 1.3 for python which can be installed using ``pip install xes'' (the edited version ``xes.py'' is available at \cite{xes}) as well as the file ``filepaths.txt''given at \cite{scenario1_filepaths} for scenario 1 and \cite{scenario2_filepaths} for scenario 2 which point to the locations of the logs that should be included have to be present:

\lstset{language=Python}
\begin{lstlisting}
import yaml;
import xes;
import os;

id_uuid = {}


with open("filepaths.txt") as f:
	filepaths = f.readlines()
filepaths = [x.strip() for x in filepaths] 
n=1
filedata = {filepath: open(filepath, 'r') for filepath in filepaths}
log = xes.Log()
log_set = False
for file in filedata.values():
	trees = yaml.load_all(file);
	t = xes.Trace()
	for tree in trees:
		if 'log' in tree:
			if log_set == False:
				log.extensions = [
					xes.Extension(name="Time",prefix="time",uri=tree['log']['extension']['time']),
					xes.Extension(name="Concept",prefix="concept",uri=tree['log']['extension']['concept']),
					xes.Extension(name="Organizational",prefix="org",uri=tree['log']['extension']['organisational']),
					xes.Extension(name="Lifecycle",prefix="lifecycle",uri=tree['log']['extension']['lifecycle'])
				]
				log.global_trace_attributes = [
					xes.Attribute(type="string",key='concept:name',value=tree['log']['global']['trace']['concept:name']),
					xes.Attribute(type="string",key='cpee:name',value=tree['log']['global']['trace']['cpee:name'])
				]
				log.global_event_attributes = [
					xes.Attribute(type="string",key='concept:name',value=tree['log']['global']['event']['concept:name']),
					xes.Attribute(type="string",key='cpee:endpoint',value=tree['log']['global']['event']['concept:endpoint']),
					xes.Attribute(type="string",key='id:id',value=tree['log']['global']['event']['id:id']),
					xes.Attribute(type="string",key='lifecycle:transition',value=tree['log']['global']['event']['lifecycle:transition']),
					xes.Attribute(type="string",key='cpee:lifecycle:transition',value=tree['log']['global']['event']['cpee:lifecycle:transition']),
					xes.Attribute(type="date",key='time:timestamp',value="1990-02-17T09:45:00.000+01:00")
				]
				log.classifiers = [
					xes.Classifier(name="Event ID Transition Classifier",keys="id:id lifecycle:transition"),
					xes.Classifier(name="MXML Legacy Classifier",keys="concept:name lifecycle:transition"),
					xes.Classifier(name="Event Name",keys="concept:name"),
					xes.Classifier(name="Event ID",keys="id:id"),
					xes.Classifier(name="CPEE Classifier",keys="concept:name cpee:lifecycle:transition"),
					xes.Classifier(name="CPEE Endpoint",keys="cpee:endpoint cpee:lifecycle:transition")
				]
				log_set= True
			t.attributes = [
				xes.Attribute(type="string", key="concept:name", value=tree['log']['trace']['concept:name']),
				xes.Attribute(type="string", key="cpee:name", value=tree['log']['trace']['cpee:name']),
				xes.Attribute(type="string", key="cpee:uuid", value=tree['log']['trace']['cpee:uuid'])
			]
		if 'event' in tree:
			if tree['event']['cpee:lifecycle:transition'] == 'activity/calling' or tree['event']['cpee:lifecycle:transition'] == 'activity/done':
				e = xes.Event()
				endpoint=""
				if('concept:endpoint' in tree['event']):
					endpoint = tree['event']['concept:endpoint']
				e.attributes = [
					xes.Attribute(type="string", key="concept:name", value=tree['event']['concept:name']),
					xes.Attribute(type="string", key="cpee:endpoint", value=endpoint),
					xes.Attribute(type="string", key="id:id", value=tree['event']['id:id']),
					xes.Attribute(type="string", key="lifecycle:transition", value=tree['event']['lifecycle:transition']),
					xes.Attribute(type="string", key="cpee:lifecycle:transition", value=tree['event']['cpee:lifecycle:transition']),
					xes.Attribute(type="date", key="time:timestamp", value=tree['event']['time:timestamp'])
				]
				t.add_event(e)
	log.add_trace(t)
	print(str(n) + " of " + str(len(filepaths)) + " logs parsed.")
	n=n+1


open("logs.xes", "w").write(str(log))
\end{lstlisting}

\subsection{Fitness}
Fitness determines how well a process log is described by the given process template. The fitness value is between 0 and 1. A value of 1 shows that a log is completely conformant to the template while lower values indicate that the execution of the log is not covered well by the template. Different kinds of fitness values exist. The ones used for this work are:

\begin{itemize}
  \item \textbf{Move-Model Fitness} is the value of correct model moves divided by the overall number of model moves during the replay.
  \item \textbf{Move-Log Fitness} is the value of correct log moves divided by the overall number of log moves during the replay.
  \item \textbf{Trace Fitness} is the cost-based fitness value as described in \cite[p.7]{replaylogonpetrinet}. This is a value obtained by subtracting the raw fitness cost divided by the maximum fitness cost from 1.
\end{itemize}

Executing the conformance check as provided in the ``Replay a Log on Petri Net for Conformance Analysis'' from the ``PNetReplayer'' ProM package on the petri net created following the rules given in section \ref{sec:confmodelling} (and creating final markings as well as making helper transitions invisible) and the XES log with standard parameters (which means choosing the option ``penalize improper completion'', choosing ``A* Cost-based Fitness Express with ILP'' as the algorithm, and setting all ``Move on Model Costs'' and ``Move on Log Costs'' to 1) makes it possible to obtain a CSV file with the abovementioned fitness values. These are used to create the tables given in section \ref{sec:confscenario1} and \ref{sec:confscenario2} showing the results. They contain one line for each log trace (or group of log traces if there are identical ones). The columns represent one of the three fitness values given above (for each of the templates). The maximum value for each line is highlighted. (green for move-model fitness, yellow for move-log fitness and red for trace fitness).

\subsection{Application to Log Data - Scenario 1}
\label{sec:confscenario1}

For the first scenario, the events from the XES log are mapped to the petri net transitions using the labels of the tasks (e.g.  Detect Lowerhousing Production Start) combined with the information if it is starting or ending.

\begin{figure}[h]
  \includegraphics[width=\textwidth]{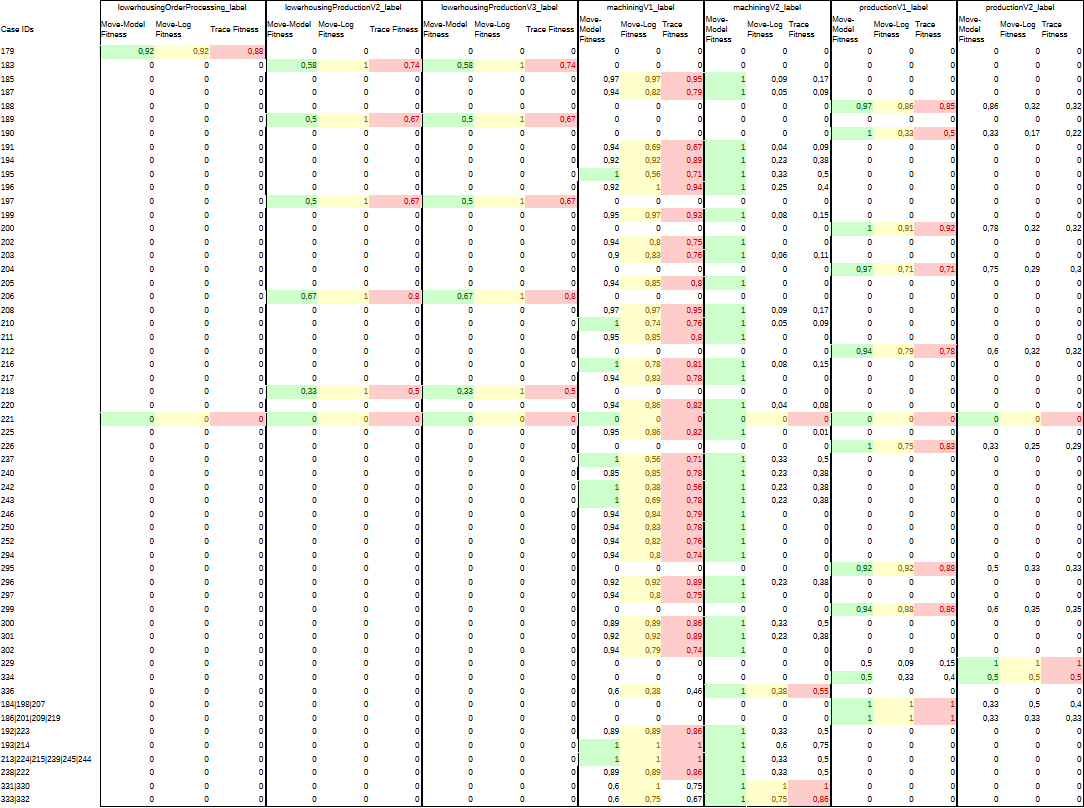}
  \caption{Conformance Checking Results Using Labels For Event/Transition Mapping}
  \label{fig:confresult1}
\end{figure}

The result table given in Fig. \ref{fig:confresult1} shows for each log how well it fits each process template and highlights the maximum value of each characteristic by giving the corresponding cell a green (maximum move-model fitness), yellow (maximum move-log fitness), or red (maximum trace fitness) background color. Based on this, one can decide an instance of which template the log is - sometimes the template of the maximum value of move-model fitness differs from the template of the maximum value of the other two characteristics, this seems to be based on a combination of short templates (e.g. MachiningV2) combined with longer logs (e.g. log of MachiningV1). If this happens, the model is moved in a correct way going through the short template while skipping many log events (which is not important for the move-model fitness) - for the longer MachiningV1 template which the trace is actually based on there is a small number of steps where a task in the model is skipped because some events might be in the wrong order or missing which leads to a slightly lower move-model fitness. These effects do not occur in the move-log fitness and trace fitness so it is concluded that these two features should be the ones to base the decision on (conveniently for all traces these two do not contradict each other when deciding which template a trace is based on so it is not necessary to think further about which of them is more significant). The results of this analysis match the ones of the manual analysis (given in templates.xlsx - available at \cite{scenario1_templates}). Therefore it is concluded that the goal of identifying the underlying template of a log using conformance checking is working as expected.

\subsection{Application to Log Data - Scenario 2}
\label{sec:confscenario2}
In addition to the approach used in the first scenario described in section \ref{sec:confscenario1}, not only the labels were used for mapping events in the log to tasks in the petri net. The two characteristics which were used to perform the abovementioned mapping for the second scenario are:
\begin{itemize}
  \item using labels (e.g. ``Detect GV12 Production Start'') as for the first scenario
  \item using endpoints (e.g. ``https://centurio.work/flow/start/url/'')
\end{itemize}

\begin{figure}[h]
  \includegraphics[width=\textwidth]{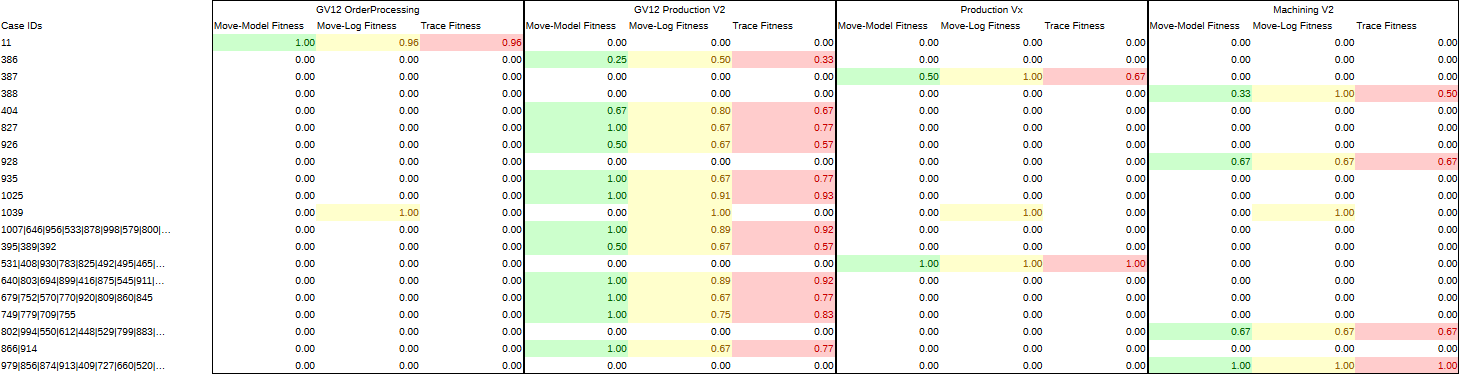}
  \caption{Conformance Checking Results Using Labels For Event/Transition Mapping}
  \label{fig:confresult2_labels}
  \includegraphics[width=\textwidth]{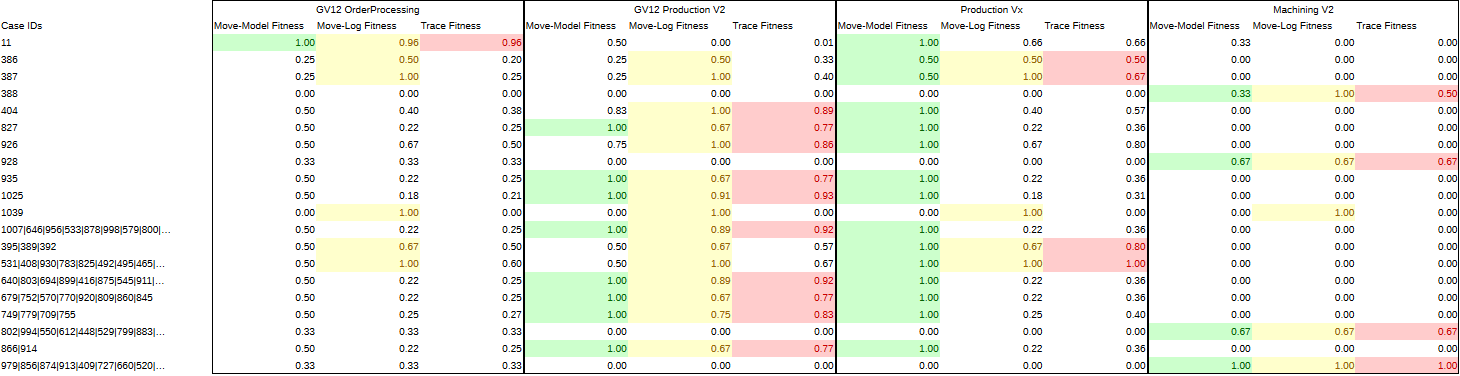}
  \caption{Conformance Checking Results Using Endpoints For Event/Transition Mapping}
  \label{fig:confresult2_endpoints}
\end{figure}

Using labels leads to a good solution in the first scenario. Using endpoints is an alternative characteristic that can be used. Therefore, the results of both characteristics are compared to each other to show the advantages and disadvantages.

As can be seen in the results given in Fig. \ref{fig:confresult2_labels} and \ref{fig:confresult2_endpoints}, the matching of the logs to the corresponding templates is similar for using labels and endpoints as characteristic for the mapping of events to petri net transitions. The following advantages are present when using labels
\begin{itemize}
  \item all logs (apart from the plain instance - log number 1039) are correctly assigned to the corresponding templates
  \item the results obtained are clear (i.e. the fitness values of a log for all but the corresponding template is 0)
\end{itemize}
and endpoints respectively
\begin{itemize}
  \item the results are more stable (i.e. the result is not dependent on the label of the tasks)
  \item endpoints describe actual functionality of a task
\end{itemize}
Clearly, there are also some disadvantages for both methods. For using labels these are
\begin{itemize}
  \item unstable if labels are changed (e.g. ``QC Shop Floor'' is renamed to ``Manually Measure'' at some point)
  \item labels are not linked to the functionality of a task
\end{itemize}
and for endpoints:
\begin{itemize}
  \item if endpoints are used, it is oftentimes difficult to determine the process template (e.g. ``start/url'' as generic template is a problem)
  \item the results are less clear (higher fitness for false templates) and sometimes even the false template is chosen
\end{itemize}

\subsection{Lessons Learned}
Overall, as the endpoints provide a much more stable result which is also based on the functionality of tasks, this would be the desired characteristic. But there is also a major drawback using it which comes from the endpoints being generic and therefore being used in many different process models (see for example the ``start/url'' endpoint). In order to tackle this problem, one could think of including the parameters of these endpoints to get a more detailed idea of the functionality performed. Unluckily, this is more difficult as it seems at first because there is no distinction between parameters for a call to an endpoint that are fixed and ones that can (at least theoretically) be changed at runtime (i.e. by using a data element which is obtained during the process execution) and therefore the usage of the information given in the template models may not always be the one which is also used at execution time. Consequently, using the value of data elements (and therefore also parameters of endpoints) is not possible because they may not be defined in the template at all but at a later point during the execution.


\section{Classification and Clustering of Log Data} 
\label{sec:class}

\subsection{Extracting Machining Data From YAML Logs}
\label{sec:clussdatapreparation}

Since the log data is available in the YAML format, several steps have to be taken to transform them into data which can be used for classification and clustering. In order to achieve this, the logs are first transformed into CSV files taking only the ``Fetch'' tasks with the ``activity/receiving'' lifecycle transition into consideration as these are the only ones containing machining data. The data extracted from the logs contains the following information that is used in the CSV files: ``source'', ``name'', ``description'', ``path'', ``value'', ``timestamp'', ``StatusCode'', ``ServerTimestamp'', ``VariantType'', ``ClientHandle''. Afterwards, it has to be defined which logs failed and which ones were successful. This differs for the two scenarios. In the first one the goal is to find out if the machining log is the last one spawned by the parent ``Production'' process which means that it is the one finishing successfully. For the second scenario successful logs are the ones which pass the measurements (the code for extracting the measurements from the YAML logs is available at \cite{scenario2_measuring}) - different measurements are used for classification as described in section \ref{sec:clussscenario2}. Using the information from the CSV files and the definition of successful logs, the data is imported into R. In order to represent all logs in the same way, some common parameters need to be chosen. This is done by only using logs which are long enough, finding parameters which occur in all of them and then choosing the ones which occur often enough for meaningful analysis. More details about the selection of logs and parameters are given in the sections where the application of the methods to the data is described (\ref{sec:clussscenario1} and \ref{sec:clussscenario2}).

The python code for extracting the machining data from the YAML log files and write them into a CSV file is given below. It is also provided at \cite{scenario2_csv_map} for the second scenario (at \cite{scenario1_csv_map} for the first scenario) and executed using the command ``python csv\_map.py'' which needs the file ``machiningFilepaths.txt'' (available at \cite{scenario2_machining_filepaths}) or for the first scenario ``filepaths\_machining.txt'' (available at \cite{scenario1_machining_filepaths}) to identify the machining logs containing the data.

\lstset{language=Python}
\begin{lstlisting}
import yaml;
import os;
import csv;
import re;
import sys;

csvData = [["Id","source","name","description","path","value","timestamp","StatusCode","ServerTimestamp","VariantType","ClientHandle"]]

with open("machiningFilepaths.txt") as f:
	filepaths = f.readlines()
filepaths = [x.strip() for x in filepaths] 
n=1
#filepaths = ['logs/production/acc7d2e4-f949-4e9b-a99a-afe3469cbbe9.xes.yaml']

filedata = {filepath: open(filepath, 'r') for filepath in filepaths}

#print(filedata)

for file in filedata.values():
	print(file.name)
	logname=str(n)
	trees = yaml.load_all(file);
	for tree in trees:
		if 'log' in tree:
				logname=tree['log']['trace']['concept:name']
		if 'event' in tree:
			if tree['event']['cpee:lifecycle:transition'] == 'activity/receiving' and tree['event']['concept:name'] == 'Fetch':
				if 'list' in tree['event']:
					for data_r in tree['event']['list']['data_receiver']:
						for data in data_r['data']:
							id=""
							source=""
							name=""
							description=""
							path=""
							value=""
							timestamp=""
							statusCode=""
							serverTimestamp=""
							variantType=""
							clientHandle=""
							if 'ID' in data:
								id=data['ID']
							if 'source' in data:
								source=data['source']
							if 'name' in data:
								name=data['name']
							if 'description' in data:
								description=data['description']
							if 'path' in data:
								path=data['path']
							if 'value' in data:
								value=str(data['value']).rstrip()
							if 'timestamp' in data:
								timestamp=data['timestamp']
							if 'meta' in data:
								if 'StatusCode' in data['meta']:
									statusCode=data['meta']['StatusCode']
								if 'ServerTimestamp' in data['meta']:
									serverTimestamp=data['meta']['ServerTimestamp']
								if 'VariantType' in data['meta']:
									variantType=data['meta']['VariantType']
								if 'ClientHandle' in data['meta']:
									clientHandle=data['meta']['ClientHandle']
							csvNewData = [id,source,name,description,path,value,timestamp,statusCode,serverTimestamp,variantType,clientHandle] 
							csvData.append(csvNewData)
				else:
					csvNewData = ["","","","","","",tree['event']['time:timestamp'],"","","",""] 
					csvData.append(csvNewData)
	print(str(n) + " of " + str(len(filepaths)) + " logs parsed.")
	csvString=""
	for v in csvData:
		line='*'.join(str(r) for r in v)
		csvString+=line+"\n"
	open("log"+str(logname)+".csv", "w").write(csvString)
	csvData=[["Id","source","name","description","path","value","timestamp","StatusCode","ServerTimestamp","VariantType","ClientHandle"]];
	n=n+1

sys.exit(0);
\end{lstlisting}

\subsection{Hierarchical Clustering}
\label{sec:clusshierarchicalclustering}
Hierarchical clustering is a method of building clusters by either starting with each observation in its own cluster or with all observations in one cluster and then merge the ones being closest to each other using some distance metric (or split into a number of clusters based on the distance in case one big cluster is used as starting point).
In order to cluster the data, only the features derived from the logs as described earlier in section \ref{sec:clussdatapreparation} were used without looking at if the log is successful or not. In order to find the appropriate number of clusters, a scree plot is created and clustering is then performed for promising cluster numbers. Additionally, a silhouette plot is created so that statements about the quality of clustering can be made. This method is only performed for the first scenario.

\subsection{K-Means Clustering}
K-Means clustering is a method where the number of clusters is defined before the algorithm is performed. The cluster centers are then randomly assigned and shift iteratively based on the mean of the contained points.
It is performed with the same data set as in section \ref{sec:clusshierarchicalclustering} using the same number of clusters as for the hierarchical clustering to check if the usage of different clustering methods has any effect on the result. A silhouette plot is also created for each clustering with different numbers of clusters. As for hierarchical clustering, this method is only performed for the first scenario.

\subsection{Feature Selection with Random Forest}
\label{sec:clussfeatureselection}
Classifying data sets with a lot of feature variables in relation to the amount of data points, comes with the need to reduce this amount of feature variables. This reduces the amount of overfitting and makes the training of data and interpretation of results easier (\cite{bermingham2015application}) which is relevant for the data sets of both scenarios, especially the first one. Goal of the feature selection is to find out which feature variables are the most important, meaning which contribute the most to the corresponding class of the data.

For the feature selection an automatic method provided by the “caret” package in R is used, namely recursive feature elimination. This algorithm builds many models with the given classification algorithm and different subsets of feature variables and compares the accuracy (\cite{caretfeatureelimination}). In this case the random forest algorithm is used to compare the models with each other. The random forest algorithm works by creating many randomly generated decision trees. Each decision tree of this forest is used for classifying the given data. The class which gets the majority of predictions of the trees is used as the final classification outcome of the forest. The advantage of the random forest algorithm is that it trains and evaluates very fast, especially with big amounts of data and is performing well in recognizing important feature variables (\cite{ho1995random}).

\subsection{Support Vector Machines}
Support vector machines is a method to classify data. This is done by constructing one or multiple hyperplanes which separate data points into different classes. This is the primary method of classification for the data sets given. The usage of SVM was done in R, using linear, radial, sigmoid and polynomial kernels. In order to perform classification the function ``svm'' from the R package ``e1071'' is used (for more information on this package see \cite{e1071}).

\subsection{Naive Bayes}
Naive Bayes is another method to classify data, based on probabilities. This was used as a comparison to SVM, to see if using a different classifier has significant effects on the classification. In order to perform classification the function ``naiveBayes'' from the R package ``e1071'' is used (for more information on this package see \cite{e1071}).

\subsection{Application to Log Data - Scenario 1}
\label{sec:clussscenario1}
Every log has not only differences in the number of variables measured but also in the length of the log (i.e. the number of data points) in general. Therefore, it is first analysed how many data points are contained in each log in order to find the ones which can be used for the analysis later on.

Obviously, the logs having only few data elements are very difficult to use for the analysis task and therefore only the logs which have more or equal to 100 data elements can be used. As 6 of 47 logs do not meet this prerequisite, only 41 logs are remaining after the filtering. The next task is to look at the parameters present in all files and then the minimum number of these IDs within the files that can be used (which means files which have $\geq100$ data elements) are calculated. Afterwards, it has to be decided which data points are used (the last 10 data elements of each usable variable in this case). Therefore, a log is represented in the following way: the last 10 values of the variables following variables are used:
\begin{itemize}
  \item ns=2;s=/Channel/MachineAxis/aaLoad[u1,1]
  \item ns=2;s=/Channel/MachineAxis/aaLoad[u1,2]
  \item ns=2;s=/Channel/MachineAxis/aaLoad[u1,3]
  \item ns=2;s=/Channel/MachineAxis/aaTorque[u1,1]
  \item ns=2;s=/Channel/MachineAxis/aaTorque[u1,2]
  \item ns=2;s=/Channel/MachineAxis/aaTorque[u1,3]
  \item ns=2;s=/Channel/Spindle/driveLoad
\end{itemize}

Therefore, a log is represented by 70 ($7*10$) parameters (+1 if the success or no success flag is also taken into consideration). This data is then used to perform the data analysis.\\
Since the amount of feature variables in the first data set is really high with 71 in comparison to the amount of data logs with 41, the risk of overfitting the data when trying to classify them has to be considered (for example using SVM as algorithm for classifying the data, the accuracy for the test and training data is, depending on the seed, nearly always or close to 100\%, which is probably an unrealistic value and that model could not be used for other data). Therefore, before classification can be applied, features must be selected. This is done using the method described in section \ref{sec:clussfeatureselection}.

\begin{figure}[h]
  \includegraphics[width=\textwidth]{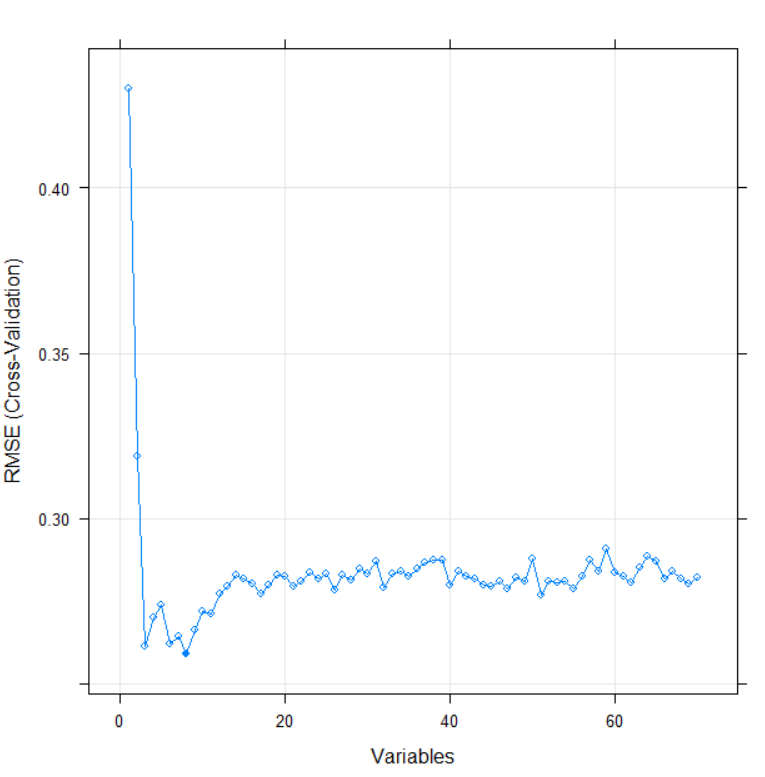}
  \caption{Recursive Feature Elimination}
  \label{fig:clussrecursivefeaturelimination}
\end{figure}

In order to get the results presented below the R code given at \cite{scenario1_classification} has to be executed. Fig. \ref{fig:clussrecursivefeaturelimination} created with this code, shows the root-mean-squared-error in comparison to the number of variables used in the model. In this case using 8 variables gives the lowest root-mean-squared-error and therefore is a good indicator, which and how many feature variables are sufficient, to create a model out of it. The variables used are (the number after the dot indicates which of the values within the last 10 - this means that ``.1'' is the 10th value and ``.10'' is the first value counting from the back):
\begin{itemize}
  \item ns=2;s=/Channel/MachineAxis/aaLoad[u1,1].7
  \item ns=2;s=/Channel/Spindle/driveLoad.8
  \item ns=2;s=/Channel/MachineAxis/aaTorque[u1,1].7
  \item ns=2;s=/Channel/MachineAxis/aaLoad[u1,1].6
  \item ns=2;s=/Channel/Spindle/driveLoad.7
  \item ns=2;s=/Channel/MachineAxis/aaLoad[u1,1].8
  \item ns=2;s=/Channel/MachineAxis/aaLoad[u1,1].5
  \item ns=2;s=/Channel/MachineAxis/aaLoad[u1,1].10
\end{itemize}

Out of those feature variables a new data frame is created. In the next step the data is split into a training set (75\%, 31 logs) and a test set (25\%, 10 logs).

For classification different algorithms are used, to give an idea about the accuracy of classification and if the models could be viable for classifying the logs.
\begin{table}[h]
  \centering
  \begin{tabular}{|l|l|l|l|l|l|}
\hline
		      & \multicolumn{2}{|l}{} & \multicolumn{1}{l}{SVM} & \multicolumn{1}{l|}{} & Naive Bayes \\ \hline
kernel		      & linear & radial & sigmoid & polynomial & - \\ \hline
cost                  & 1            & 1            & 1             & 1                & -           \\ \hline
gamma                 & 0.125        & 0.125        & 0.125         & 0.125            & -           \\ \hline
degree                & -            & -            & -             & 3                & -           \\ \hline
coef.0                & -            & -            & 0             & 0                & -           \\ \hline
\# support            & 9            & 12           & 12            & 13               & -           \\
vectors               &              &              &               &                  &             \\ \hline
accuracy              & 96.77\%      & 93.55\%      & 93.55\%       & 87.10\%          & 77.42\%     \\
training set          &              &              &               &                  &             \\ \hline
accuracy              & 90.00\%      & 90.00\%      & 90.00\%       & 80.00\%          & 80.00\%     \\
test set              &              &              &               &                  &             \\ \hline
  \end{tabular}
  \caption{Classification Results With Different Kernels For Scenario 1}
  \label{tab:classificationresults_scenario1}
\end{table}

The accuracy given in Tab. \ref{tab:classificationresults_scenario1} using naive bayes with 77,42\% for the training set and 80\% for the test set, seems worse than the SVMs. This could indicate that SVM is a good algorithm for this data set. The accuracies for the training set with 93,55\% for both radial and sigmoid and 87,1\% for polynomial are all decently high, but still lower than SVM with a linear kernel. Furthermore, the accuracy for the test set with 90\% for both radial and sigmoid and 80\% for polynomial, are equal or a bit less than the SVM with linear kernel, but regarding the fact that there are only 10 data sets in the test set, jumps in accuracy are very high for single correct or incorrect predictions.

\begin{figure}[h]
  \begin{subfigure}[b]{0.5\textwidth}
    \includegraphics[width=\textwidth]{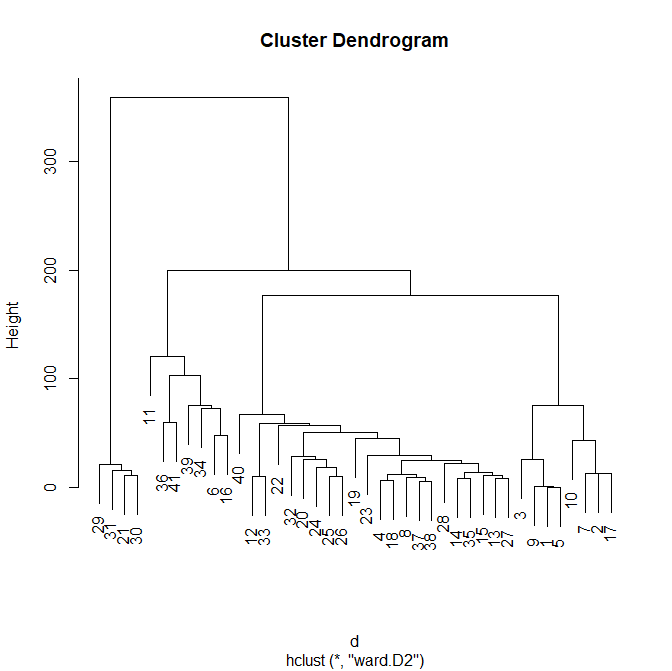}
    \subcaption{Hierarchical Clustering}
  \end{subfigure}
  \begin{subfigure}[b]{0.5\textwidth}
    \includegraphics[width=\textwidth]{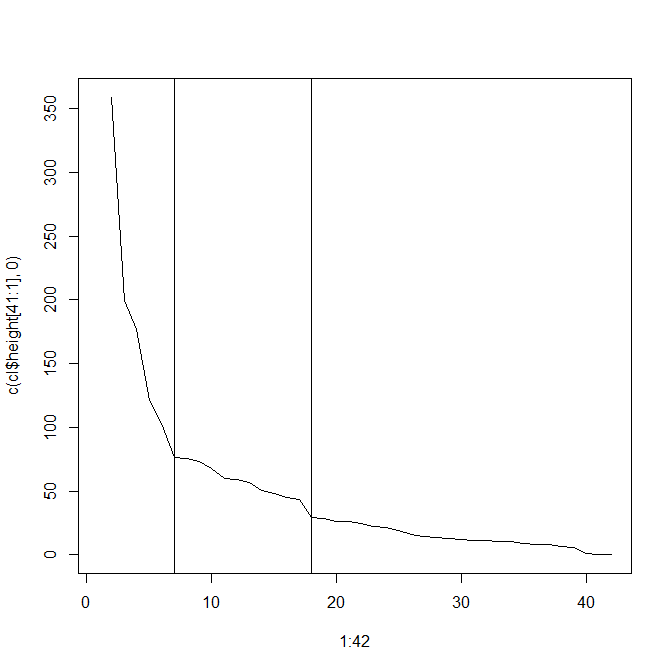}
    \subcaption{Scree Plot}
    \label{fig:clussscreeplot}
  \end{subfigure}
  \caption{Hierarchical Clustering Plots}
  \label{fig:clussclustering}
\end{figure}

In addition to classification, clustering is performed using the code given in \cite{scenario1_clusteringFullData}. The scree plot (Fig. \ref{fig:clussscreeplot}) indicates that 7 or 18 clusters should be used because these are the locations where an “elbow” can be seen. Additionally, the clustering is performed for 2 clusters because there are two cases in the original problem statement (success or not success).

\begin{figure}[h]
  \begin{subfigure}[b]{0.32\textwidth}
    \includegraphics[width=\textwidth]{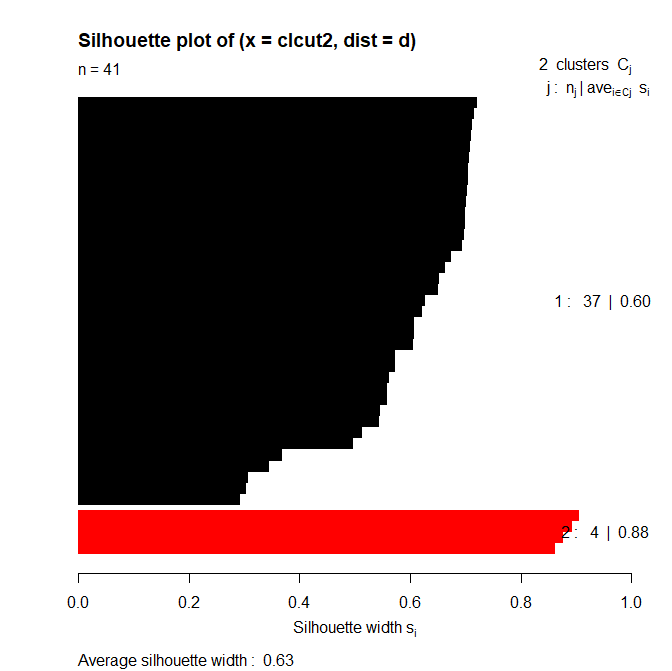}
    \subcaption{2 Clusters}
  \end{subfigure}
  \begin{subfigure}[b]{0.32\textwidth}
    \includegraphics[width=\textwidth]{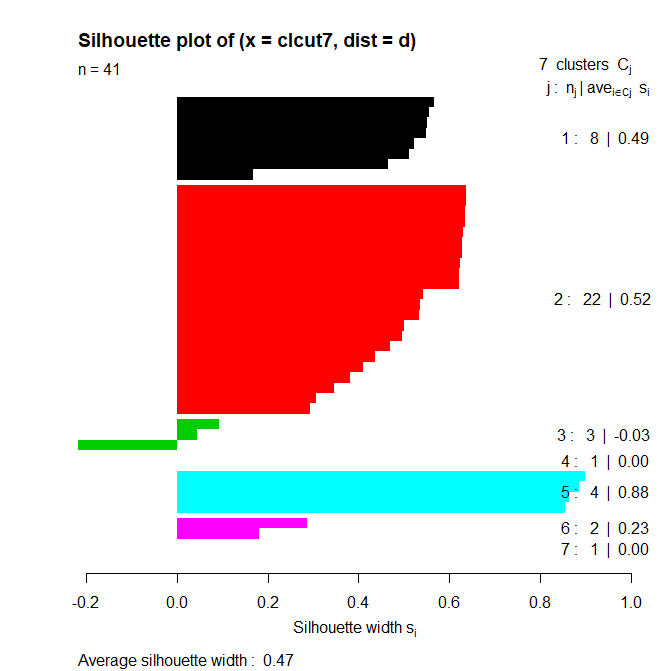}
    \subcaption{7 Clusters}
  \end{subfigure}
  \begin{subfigure}[b]{0.32\textwidth}
    \includegraphics[width=\textwidth]{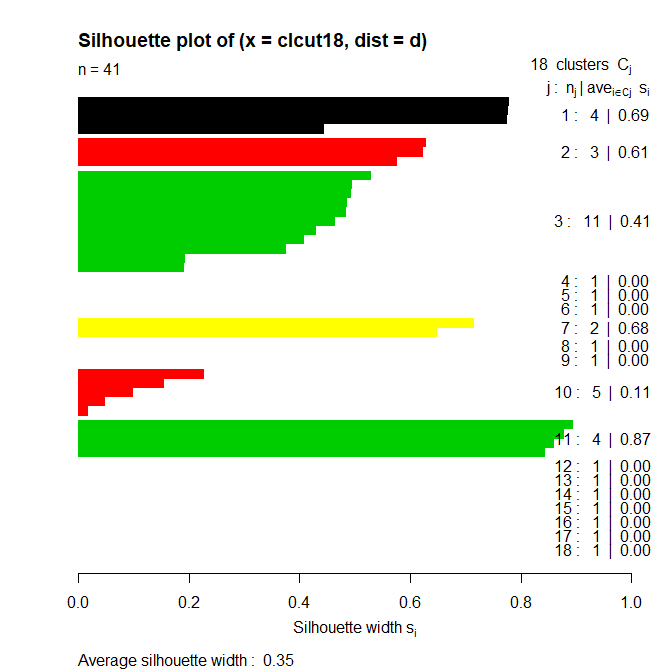}
    \subcaption{18 Clusters}
  \end{subfigure}
  \caption{Silhouette Plot For Different Numbers of Clusters Using Hierarchical Clustering}
  \label{fig:silhierarchical}
\end{figure}

The silhouette plots in Fig. \ref{fig:silhierarchical} are created for the results with different numbers of clusters.

If a class is assigned to each cluster (the one which has the highest number of points inside the cluster is chosen), an estimation of the quality of the clustering is obtained:

\begin{itemize}
  \item 2 clusters: 68.83\%
  \item 7 clusters: 92.68\%
  \item 18 clusters: 97.56\%
\end{itemize}

Obviously, the clustering using 18 clusters is the most accurate one but looking at the ratio of clusters to number of logs, and the small number of logs in most of the clusters this approach looks like overfitting. Also, the average silhouette width decreases when more clusters are used which indicates that the points are not as well located in the cluster compared to fewer clusters.

Doing clustering with the same data described above but using kmeans leads to the following results:
\begin{itemize}
  \item 2 clusters: 68.83\%
  \item 7 clusters: 92.68\%
  \item 18 clusters: 95.12\%
\end{itemize}

\begin{figure}[h]
  \begin{subfigure}[b]{0.32\textwidth}
    \includegraphics[width=\textwidth]{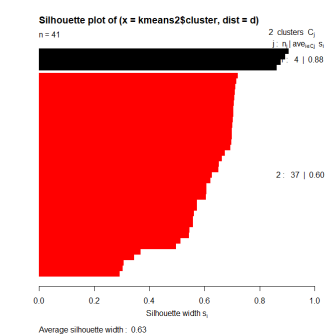}
    \subcaption{2 Clusters}
  \end{subfigure}
  \begin{subfigure}[b]{0.32\textwidth}
    \includegraphics[width=\textwidth]{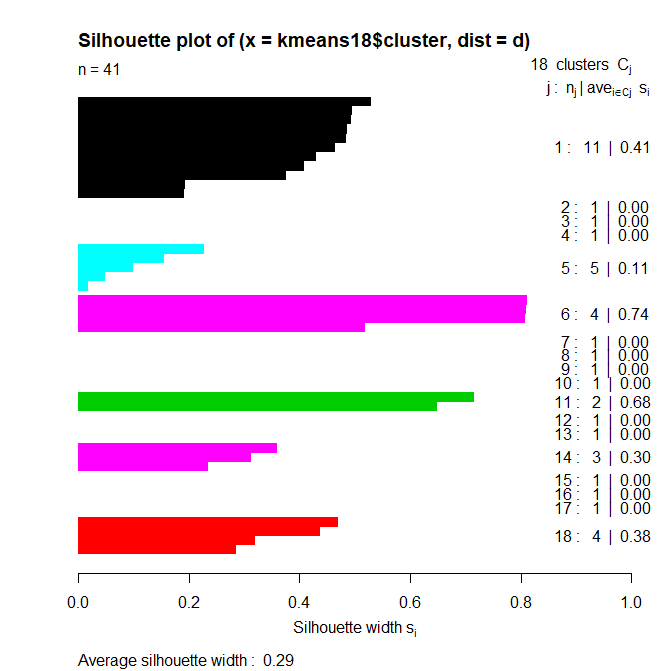}
    \subcaption{7 Clusters}
  \end{subfigure}
  \begin{subfigure}[b]{0.32\textwidth}
    \includegraphics[width=\textwidth]{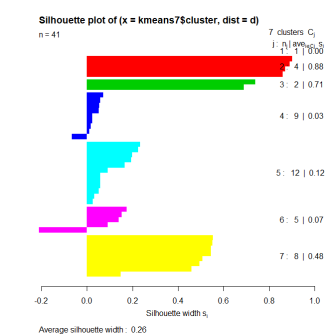}
    \subcaption{18 Clusters}
  \end{subfigure}
  \caption{Silhouette Plot For Different Numbers of Clusters Using K-Means Clustering}
\end{figure}

The results are very similar to the ones above concerning accuracy therefore again overfitting for 18 clusters (and maybe even 7) has to be taken into account. A notable difference to the hierarchical clustering is that the average silhouette width for 7 and 18 clusters is lower.

\begin{figure}[h]
  \begin{subfigure}[b]{0.5\textwidth}
    \includegraphics[width=\textwidth]{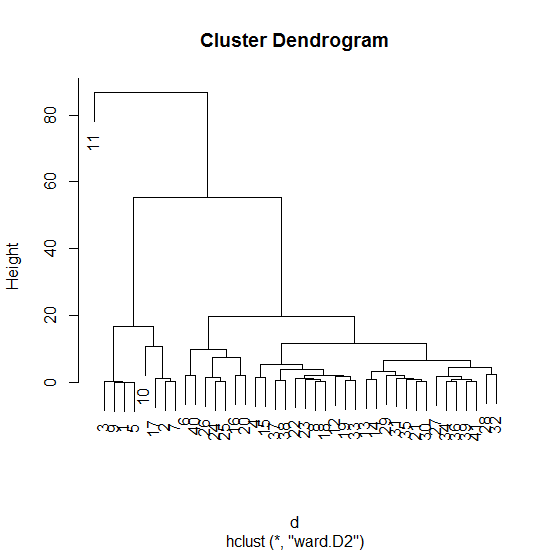}
    \subcaption{Hierarchical Clustering With Less Feature Variables}
  \end{subfigure}
  \begin{subfigure}[b]{0.5\textwidth}
    \includegraphics[width=\textwidth]{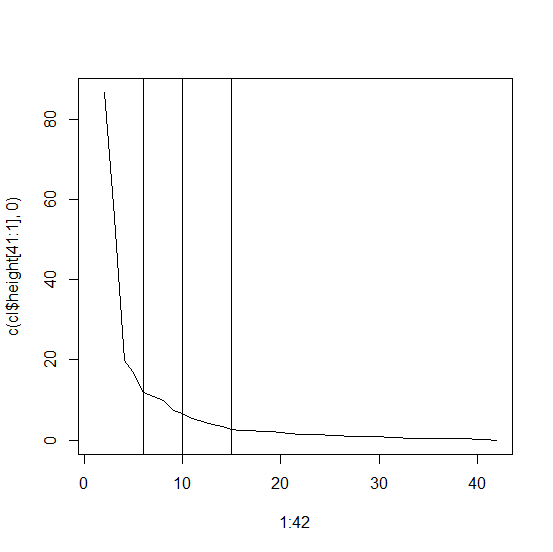}
    \subcaption{Scree Plot With Less Feature Variables}
    \label{fig:clussscreeplotsmalldataset}
  \end{subfigure}
  \caption{Hierarchical Clustering With Less Feature Variables Plots}
  \label{fig:clussclusteringsmalldataset}
\end{figure}

After performing the clustering with the data described above, the question arised how clustering would look if less features are used – therefore clustering is done again this time not using the last 10 values of each ``good'' parameter but only the first one of them (i.e. the 10th value counting from the last value). The results gathered when executing the R code available at \cite{scenario1_clusteringLessData} are given in Fig. \ref{fig:clussclusteringsmalldataset}.

The plots given in Fig. \ref{fig:clussclusteringsmalldataset} are similar to the plots for the full data set given in Fig. \ref{fig:clussclustering} - the scree plot (see Fig. \ref{fig:clussscreeplotsmalldataset}) shows which number of clusters should be used in further analysis.

Using hierarchical clustering, the following accouracies are derived:
\begin{itemize}
  \item 2 clusters: 58.54\%
  \item 6 clusters: 82.93\%
  \item 10 clusters: 82.93\%
  \item 15 clusters: 90.24\%
\end{itemize}

\begin{figure}[h]
  \begin{subfigure}[b]{0.24\textwidth}
    \includegraphics[width=\textwidth]{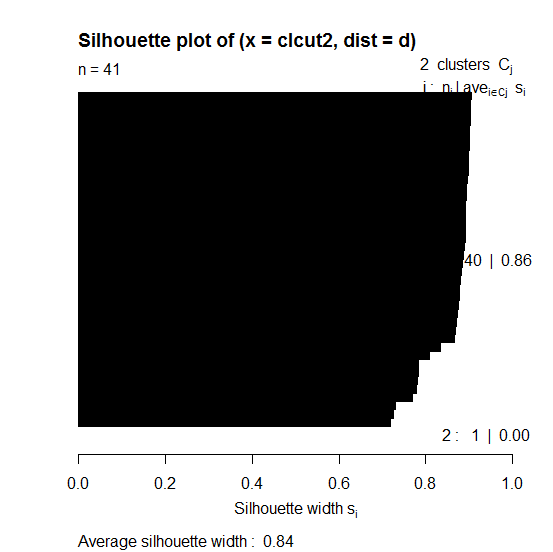}
    \subcaption{2 Clusters}
  \end{subfigure}
  \begin{subfigure}[b]{0.24\textwidth}
    \includegraphics[width=\textwidth]{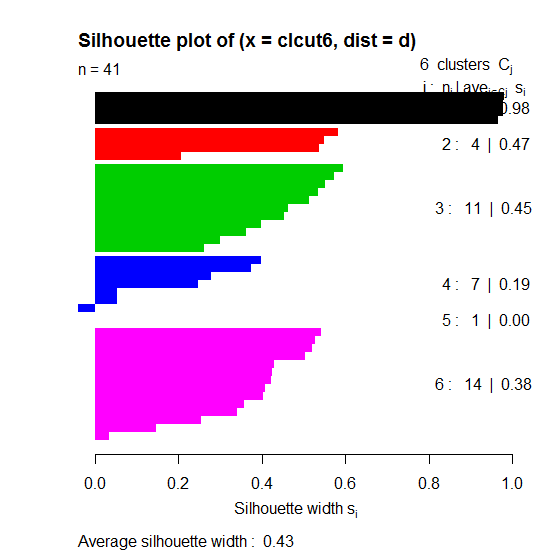}
    \subcaption{6 Clusters}
  \end{subfigure}
  \begin{subfigure}[b]{0.24\textwidth}
    \includegraphics[width=\textwidth]{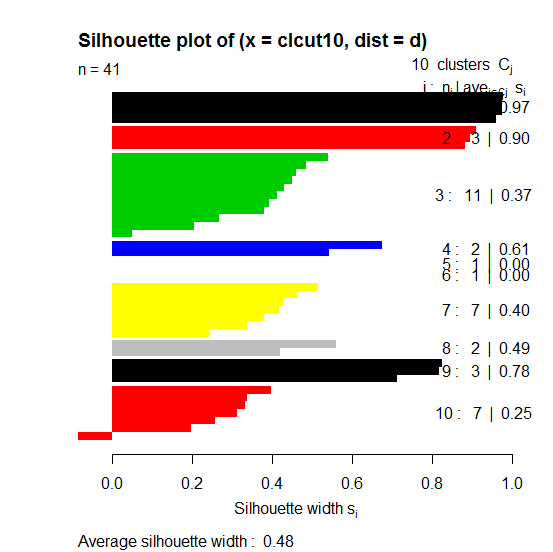}
    \subcaption{10 Clusters}
  \end{subfigure}
  \begin{subfigure}[b]{0.24\textwidth}
    \includegraphics[width=\textwidth]{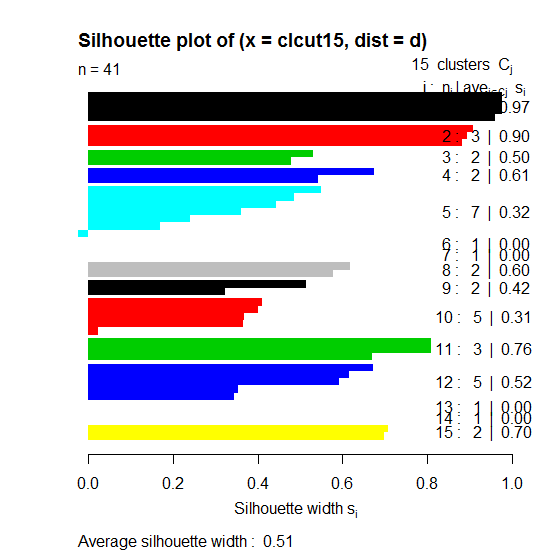}
    \subcaption{15 Clusters}
  \end{subfigure}
  \caption{Silhouette Plot For Different Numbers of Clusters Using Hierarchical Clustering With Less Feature Variables}
\end{figure}

For k-means clustering the result is as follows:
\begin{itemize}
  \item 2 clusters: 75.61\%
  \item 6 clusters: 80.49\%
  \item 10 clusters: 80.49\%
  \item 15 clusters: 87.80\%
\end{itemize}

\begin{figure}[h]
  \begin{subfigure}[b]{0.24\textwidth}
    \includegraphics[width=\textwidth]{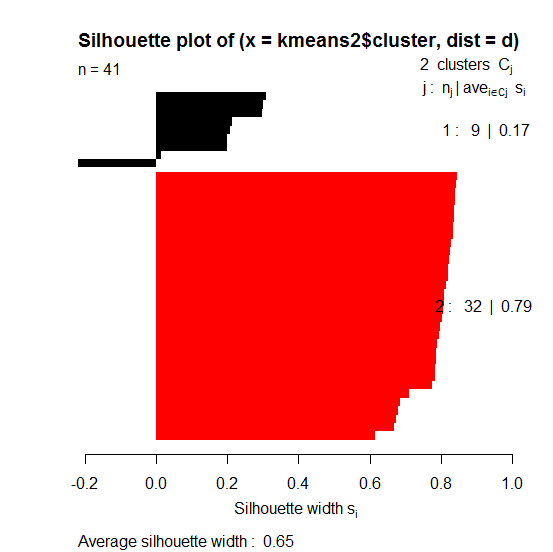}
    \subcaption{2 Clusters}
  \end{subfigure}
  \begin{subfigure}[b]{0.24\textwidth}
    \includegraphics[width=\textwidth]{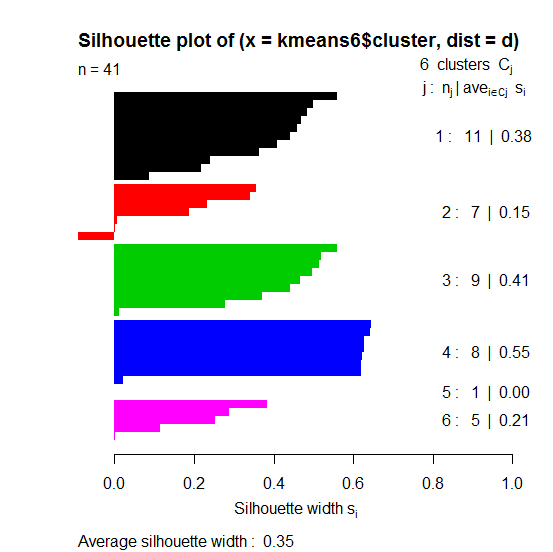}
    \subcaption{6 Clusters}
  \end{subfigure}
  \begin{subfigure}[b]{0.24\textwidth}
    \includegraphics[width=\textwidth]{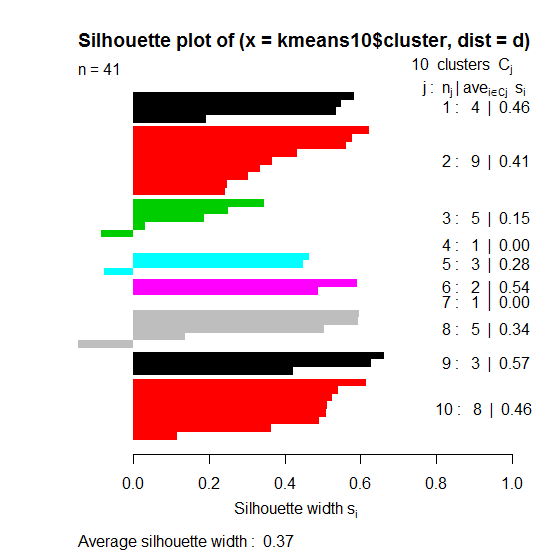}
    \subcaption{10 Clusters}
  \end{subfigure}
  \begin{subfigure}[b]{0.24\textwidth}
    \includegraphics[width=\textwidth]{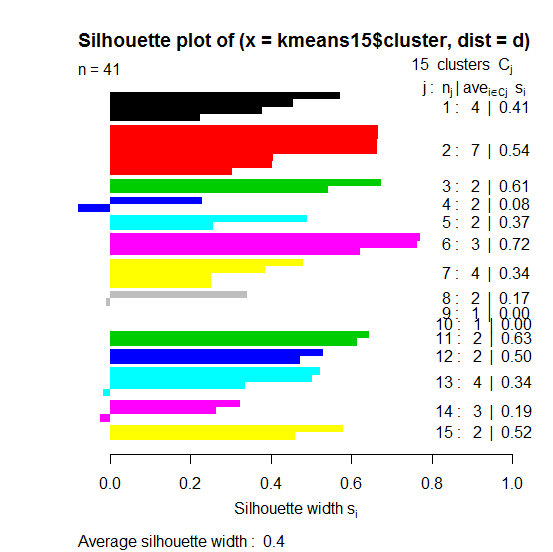}
    \subcaption{15 Clusters}
  \end{subfigure}
  \caption{Silhouette Plot For Different Numbers of Clusters Using K-Means Clustering With Less Feature Variables}
\end{figure}

The accuracies achieved with a lower number of features are slightly lower than the ones achieved with a higher number of features – but they are still not too bad. Probably a number of features between the two would be optimal but a more in-depth analysis only seems feasible if an analysis of the effect the number of logs has on the results (i.e. does the approach even work for a larger number of logs) has been done.

\subsection{Application to Log Data - Scenario 2}
\label{sec:clussscenario2}

To start working with the second data set it has to be determined which of the parameters can be used for representing a log and ultimately be used in classification. Moreover, the logs have to be checked in order to find out which of them might not be appropriate for usage because they are somehow damaged or simply too short.

First off, it is analysed how many data points are contained in each log. Afterwards the decision that only logs which contain more than 100 data points are feasible to use is established. Therefore, 3 of the 205 logs are not taken into account for the further steps meaning that 202 logs can still be used for the analysis. Using the 202 remaining logs, all feature variables which are present in all of them are determined. From these 20 parameters the minimum number of occurrence within the 202 valid logs is determined. A threshold of 10 is chosen and all of the feature variables which are below that are deleted, which leads to 14 parameters. The code used for this process is given at \cite{scenario2_analysis}.

As there are about 70 parameters representing a log in the first scenario (10 last values of 7 IDs), which achieve feasible results, it is concluded that a similar number should be chosen. Therefore, the decision that 5 data points per parameter should represent a log (as $14*5$ equals 70) is taken. For the first scenario the last 10 data points of each log are used - for this scenario it is decided to use the last 5, 5 in the middle and the first 5 data points in order to represent logs as it is interesting which of these variants provides the best results also indicating which point in time of the manufacturing might be most significant for the part to be correct. The parameters used for representation are:
\begin{itemize}
  \item ns=2;s=/Channel/MachineAxis/aaLeadP[u1,1]
  \item ns=2;s=/Channel/MachineAxis/aaLeadP[u1,2]
  \item ns=2;s=/Channel/MachineAxis/aaLeadP[u1,3]
  \item ns=2;s=/Channel/MachineAxis/aaLoad[u1,1]
  \item ns=2;s=/Channel/MachineAxis/aaLoad[u1,2]
  \item ns=2;s=/Channel/MachineAxis/aaLoad[u1,3]
  \item ns=2;s=/Channel/MachineAxis/aaTorque[u1,1]
  \item ns=2;s=/Channel/MachineAxis/aaTorque[u1,2]
  \item ns=2;s=/Channel/MachineAxis/aaTorque[u1,3]
  \item ns=2;s=/Channel/MachineAxis/aaVactB[u1,1]
  \item ns=2;s=/Channel/MachineAxis/aaVactB[u1,2]
  \item ns=2;s=/Channel/MachineAxis/aaVactB[u1,3]
  \item ns=2;s=/Channel/Spindle/actSpeed
  \item ns=2;s=/Channel/Spindle/driveLoad
\end{itemize}

The code for reading the data from the CSV files and preparing it for usage in R is given below and at \cite{scenario2_daten_einlesen_ende}. It is already taken into account that the measurements are missing for 6 of the parts - these are not used for classification later on - therefore, 196 logs are remaining.

\lstset{language=R}
\begin{lstlisting}
#-------- Daten einlesen

#install.packages("gdata")
#install.packages("cluster")
#install.packages("clue")
#install.packages("rlist")
library(gdata)
library(cluster)
library(clue)
library(rlist)

setwd("P:\\Daten\\Uni\\Master Wirtschaftsinformatik\\2018_WS\\VU Business Intelligence II\\aufgabe4\\frage2")
#setwd("C:\\Users\\Admin\\Desktop\\bi2")

files<-list.files("machining")
files<-append(files[-c(1:10)],files[1:10])

alldata<-list()

for(i in 1:length(files)) {
  alldata[[i]]<-read.csv(paste("machining\\",files[i],sep=""),header=TRUE,sep="*",stringsAsFactors=FALSE)
}

allsplitdata<-list()
for(i in c(1:length(files))) {
  dataset<-alldata[[i]]
  ordereddata<-dataset[order(dataset$Id,dataset$ServerTimestamp,dataset$timestamp),]
  splitdata<-split(ordereddata,ordereddata$Id)
  allsplitdata[[i]]<-splitdata
}

dataitems<-array()
for(i in 1:length(files)) {
  dataitems[i]<-nrow(alldata[i][[1]])
}

goodfiles<-c()
for(i in 1:length(files)) {
  if(dataitems[i]>=100) {
    goodfiles<-c(goodfiles,i)
  }
}

listofids<-c()
for(i in goodfiles) {
  listofids<-c(listofids,names(allsplitdata[[i]]))
}

ids=unique(listofids)

sumofids<-array()
for(i in c(1:length(ids))) {
  counter<-0
  for(j in c(1:length(listofids))) {
    if(ids[i]==listofids[j]) counter<-counter+1
  }
  sumofids[i]<-counter
}

idswithsums<-cbind(ids,sumofids)

goodids<-c()
for(i in c(1:length(ids))) {
  if(idswithsums[i,2]=="202") goodids<-c(goodids,idswithsums[i,1])
}

numbers<-data.frame()
for(i in goodfiles) {
  buffer<-array()
  buffer[1]<-i
  for(j in c(1:length(goodids))) {
    splitid<-which(goodids[j] == names(allsplitdata[[i]]))
	buffer[j+1]<-nrow(allsplitdata[[i]][[splitid]])
  }
  numbers<-rbind(numbers,buffer)
}
numbers<-setNames(numbers,c("filenumber",goodids))

minimum<-array()
for(i in 2:21) {
  minimum[i-1]<-min(numbers[i])
}

idswithminimum<-cbind(goodids,minimum)

reallygoodids<-c()
for(i in c(1:length(goodids))) {
  if(as.numeric(idswithminimum[i,2])>=10) reallygoodids<-c(reallygoodids,idswithminimum[i,1])
}

gooddata<-data.frame();
for(i in goodfiles) {
  buffer<-array()
  buffer1<-array()
  buffer2<-array()
  for(j in c(1:length(reallygoodids))) {
    splitid<-which(reallygoodids[j] == names(allsplitdata[[i]]))
	#for(k in c(1:5)) {
	#  buffer[(j-1)*5+k]<-as.numeric(allsplitdata[[i]][[splitid]]$value[k])
	#}
	#for(k in c(1:5)) {
	#  buffer[(j-1)*5+k]<-as.numeric(allsplitdata[[i]][[splitid]]$value[round(nrow(allsplitdata[[i]][[splitid]])/2)-3+k])
	#}
	for(k in c(1:5)) {
	  buffer[(j-1)*5+k]<-as.numeric(allsplitdata[[i]][[splitid]]$value[nrow(allsplitdata[[i]][[splitid]])-(5-k)])
	}
  }
  gooddata<-rbind(gooddata,buffer)
}
gooddatanames=c()
for(i in c(1:length(reallygoodids))) {
  for(j in c(1:5)) {
    gooddatanames=c(gooddatanames,paste(reallygoodids[i],j,sep=""))
  }
}
gooddata<-setNames(gooddata,gooddatanames)

successdata<-read.csv("measuring.csv",header=TRUE,sep="*",stringsAsFactors=FALSE)
#successdataold<-successdata

helperMM1<-successdata$MM1[179]
helperMM2<-successdata$MM2[179]
helperMM3<-successdata$MM3[179]

helper1MM1<-successdata$MM1[180]
helper1MM2<-successdata$MM2[180]
helper1MM3<-successdata$MM3[180]

for(i in c(180:129)) {
  successdata$MM1[i]<-successdata$MM1[i-2]
  successdata$MM2[i]<-successdata$MM2[i-2]
  successdata$MM3[i]<-successdata$MM3[i-2]
}

successdata$MM1[127]<-helperMM1
successdata$MM2[127]<-helperMM2
successdata$MM3[127]<-helperMM3

successdata$MM1[128]<-helper1MM1
successdata$MM2[128]<-helper1MM2
successdata$MM3[128]<-helper1MM3


goodsuccessdata<-data.frame()
for(i in goodfiles) {
  goodsuccessdata<-rbind(goodsuccessdata,successdata[i,c(3:17)])
}
gooddata<-cbind(gooddata,goodsuccessdata)


j<-1
badgooddata<-array()
for(i in c(nrow(gooddata)):1) {
  i<-i
  if(is.na(gooddata$MM1[i]) || gooddata$MM1[i]=="") {
    badgooddata[j]<-i
    j<-j+1
  }
}


for(i in badgooddata) {
  gooddata<-gooddata[-i,]
}


for(i in c(71:85)) {
  gooddata[,c(i)]<-(as.logical(gooddata[,c(i)]))
}

gooddata<-cbind(gooddata, acceptance=rowSums(gooddata[,77:85])==9)

gooddata<-cbind(gooddata, acceptance_MM=rowSums(gooddata[,71:73])==3)
\end{lstlisting}

As described above, different quintetts of data points are used to represent the logs. The difference between using the last, first, or middle five points lies in the following lines in the code:

For using the first five data points per parameter for representation the following code is used (available at \cite{scenario2_daten_einlesen_anfang}):

\lstset{language=R, firstline=92, lastline=109}
\begin{lstlisting}
for(i in goodfiles) {
  buffer<-array()
  buffer1<-array()
  buffer2<-array()
  for(j in c(1:length(reallygoodids))) {
    splitid<-which(reallygoodids[j] == names(allsplitdata[[i]]))
	for(k in c(1:5)) {
	  buffer[(j-1)*5+k]<-as.numeric(allsplitdata[[i]][[splitid]]$value[k])
	}
	#for(k in c(1:5)) {
	#  buffer[(j-1)*5+k]<-as.numeric(allsplitdata[[i]][[splitid]]$value[round(nrow(allsplitdata[[i]][[splitid]])/2)-3+k])
	#}
	#for(k in c(1:5)) {
	#  buffer[(j-1)*5+k]<-as.numeric(allsplitdata[[i]][[splitid]]$value[nrow(allsplitdata[[i]][[splitid]])-(5-k)])
	#}
  }
  gooddata<-rbind(gooddata,buffer)
}
\end{lstlisting}

For using the middle five data points per parameter for representation the following code is used (available at \cite{scenario2_daten_einlesen_mitte}):

\lstset{language=R, firstline=92, lastline=109}
\begin{lstlisting}
for(i in goodfiles) {
  buffer<-array()
  buffer1<-array()
  buffer2<-array()
  for(j in c(1:length(reallygoodids))) {
    splitid<-which(reallygoodids[j] == names(allsplitdata[[i]]))
	#for(k in c(1:5)) {
	#  buffer[(j-1)*5+k]<-as.numeric(allsplitdata[[i]][[splitid]]$value[k])
	#}
	for(k in c(1:5)) {
	  buffer[(j-1)*5+k]<-as.numeric(allsplitdata[[i]][[splitid]]$value[round(nrow(allsplitdata[[i]][[splitid]])/2)-3+k])
	}
	#for(k in c(1:5)) {
	#  buffer[(j-1)*5+k]<-as.numeric(allsplitdata[[i]][[splitid]]$value[nrow(allsplitdata[[i]][[splitid]])-(5-k)])
	#}
  }
  gooddata<-rbind(gooddata,buffer)
}
\end{lstlisting}

Looking at the measurements/tests it is noteworthy that some of them are passed/not passed every time (or close to every time) which makes them difficult to use for classification because predicting the class which occurs very often provides a good result without having to rely on actual data.

\begin{figure}[h]
  \includegraphics[width=\textwidth]{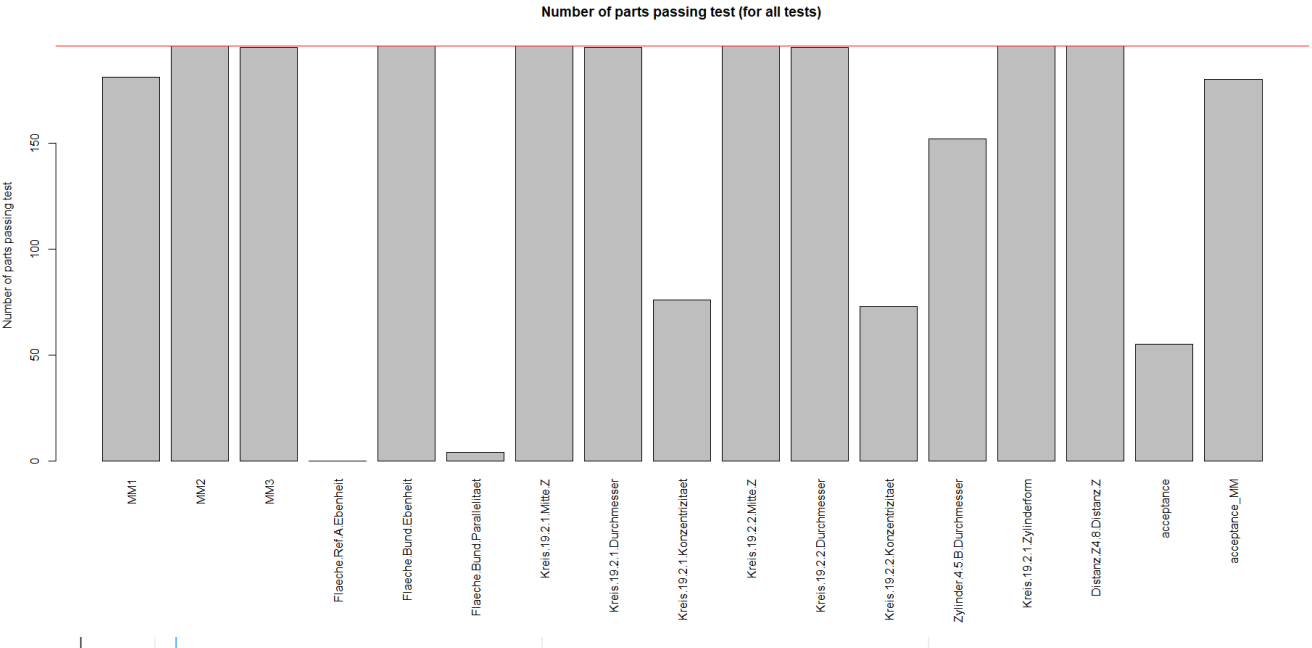}
  \caption{Number of Parts Passing Test (For All Tests)}
  \label{fig:teststats}
\end{figure}

The chart in Fig. \ref{fig:teststats} shows the number of parts passing the test for each test - the red line indicates the number of parts produced so the number of parts not passing the test can also be determined. As described above, measurements which have a good ratio of ``passed'' to ``not passed'' should be used for the classification - therefore, ``Kreis.19.2.1.Konzentrizitaet'', ``Kreis.19.2.2.Konzentrizitaet'' and ``Zylinder.4.5.B.Durchmesser'' seem good to use. In order to also have a manual measurement in the classification the best one of these (``MM1'') is also added. The two bars on the right side represent the overall acceptance for automatic (all but the first 3 automatic measurements are OK - i.e. all but the ones starting with ``Flaeche'') and manual measurements (all three manual measurements are OK).

The goal is to find out if the result of a single measurement can be predicted via classification. For the classification the last 5 data points of each machining log are used. Using the last few data points worked well for the first scenario. Using the last 5 data points of each feature variable gives 70 variables in total and 196 usable data points for each of those variables, which is a decent ratio. Measurements are chosen, where there is a good ratio between true and false (ok and nok), since trying to predict a measurement result, which is 99\% ok, is probably rather pointless. Therefore, 1 manual measurement and 3 automatic measurements are chosen, which have a good ratio. Those measurements are, including their ratio of true/false:
\begin{itemize}
\item Manual Measurement 1 (181 ok/true(92,3\%), 15 nok/false(7,7\%))
\item Kreis 19,2-1 Konzentrizitaet (76 ok/true(38,8\%), 120 nok/false(61,2\%))
\item Kreis 19,2-2 Konzentrizitaet (73 ok/true(37,2\%), 123 nok/false(62,8\%))
\item Zylinder 4,5-B Durchmesser (152 ok/true(77,6\%), 44 nok/false(22,4\%))
\end{itemize}
Although manual measurement 1 has a really high number of measurements with the result ok, it is still the lowest of all manual measurements, and it is useful to look at classification results for one manual measurement as well, to see if it behaves differently.

The data was split into a training set (75\%, 143 logs) and a test set (25\%, 53 logs).

For classification different algorithms were used, to give an idea about the accuracy of classification and if the models could be viable for classifying the logs. The code used for this analysis as well as the R output can be found at \cite{scenario2_ausgabe_r}. \\

\begin{table}[]
  \centering
  \begin{tabular}{|l|l|l|l|l|l|l|}
\hline
		      & \multicolumn{2}{|l}{} & \multicolumn{1}{l}{SVM} & \multicolumn{2}{l|}{} & Naive Bayes \\ \hline
kernel		      & linear & radial & radial (tuned) & sigmoid & polynomial & - \\ \hline
cost                  & 1            & 1            & 0.01                     & 1             & 1                & -           \\ \hline
gamma                 & 0.01428571   & 0.01428571   & 0.001                    & 0.01428571    & 0.01428571       & -           \\ \hline
degree                & -            & -            & -                        & -             & 3                & -           \\ \hline
coef.0                & -            & -            & -                        & 0             & 0                & -           \\ \hline
\# support            & 52           & 91           & 31                       & 32            & 83               & -           \\
vectors               &              &              &                          &               &                  &             \\ \hline
accuracy              & 97.90\%      & 91.61\%      & 90.91\%                  & 90.91\%       & 93.71\%          & 72.73\%     \\
training set          &              &              &                          &               &                  &             \\ \hline
accuracy              & 84.91\%      & 96.23\%      & 96.23\%                  & 96.23\%       & 94.34\%          & 54.72\%     \\
test set              &              &              &                          &               &                  &             \\ \hline
  \end{tabular}
  \caption{Classification Results For ``ManualMeasurement1'' Using The Last 5 Data Points}
  \label{tab:clussclassificationresultsMM1last}
\end{table}

Classification was first performed for ``ManualMeasurement1'' using the last 5 data points (Tab. \ref{tab:clussclassificationresultsMM1last}). The accuracy for the SVM with linear kernel with 97,9\% for the training set and 84,9\% for the test set seems very good, but looking at the predictions, which all predict only one class and the initial ratio of true/false, it can be seen that the classification is indeed very bad. The accuracies in the testsets for other kernels with 94,3\% for polynomial and 96,2\% for radial and sigmoid also seem really good, but a closer look shows, that the SVM figured out to just predict everything true, therefore getting a really high accuracy, but is basically not useful for actual predicting. Naive Bayes doesn’t work at all here.

Those high accuracies can be attributed to the fact that the initial ratio of manual measurement 1 is really skewed in favour of true/ok. Therefore classification doesn't seem viable here. In the next step classification is used for the automatic measurements, which have a far more balanced ratio between true/ok and false/nok and should, if classification is viable, provide better results.

For classification for the measurement results of Kreis 19,2-1 Konzentrizitaet, Kreis 19,2-2 Konzentrizitaet and Zylinder 4,5-B Durchmesser the same steps as above for Manual Measurement 1 are used, therefore just the results of the different SVM kernels and naive bayes are presented in the Tab. \ref{tab:clussclassificationresultsKreis1921Konzentrizitaetlast}, \ref{tab:clussclassificationresultsKreis1922Konzentrizitaetlast}, and \ref{tab:clussclassificationresultsKreisZylinder45BDurchmesserlast}.

\begin{table}[]
  \centering
  \begin{tabular}{|l|l|l|l|l|l|l|}
\hline
		      & \multicolumn{2}{|l}{} & \multicolumn{1}{l}{SVM} & \multicolumn{2}{l|}{} & Naive Bayes \\ \hline
kernel		      & linear & radial & radial (tuned) & sigmoid & polynomial & - \\ \hline
cost                  & 1            & 1            & 1                        & 1             & 1                & -           \\ \hline
gamma                 & 0.01428571   & 0.01428571   & 0.1                      & 0.01428571    & 0.01428571       & -           \\ \hline
degree                & -            & -            & -                        & -             & 3                & -           \\ \hline
coef.0                & -            & -            & -                        & 0             & 0                & -           \\ \hline
\# support            & 103          & 129          & 143                      & 117           & 132              & -           \\
vectors               &              &              &                          &               &                  &             \\ \hline
accuracy              & 82.52\%      & 77.62\%      & 98.60\%                  & 58.74\%       & 74.13\%          & 66.43\%     \\
training set          &              &              &                          &               &                  &             \\ \hline
accuracy              & 56.60\%      & 66.04\%      & 66.04\%                  & 62.26\%       & 62.26\%          & 52.83\%     \\
test set              &              &              &                          &               &                  &             \\ \hline
  \end{tabular}
  \caption{Classification Results For ``Kreis19,2-1 Konzentrizitaet'' Using The Last 5 Data Points}
  \label{tab:clussclassificationresultsKreis1921Konzentrizitaetlast}
\end{table}

The results for Kreis19,2-1 Konzentrizitaet (Tab. \ref{tab:clussclassificationresultsKreis1921Konzentrizitaetlast}) show that the accuracies of the testsets of the the different classification methods range from 52,8\% (Naive Bayes) to 66\% (SVM with radial kernel). This is not a good accuracy at all and it seems like the classification does not really find a good way to predict the outcome. Apparently the SVM found its best way to predict, by predicting nearly everything as the label which has the majority of the dataset (the data set has about 62\% labelled false). As can be seen for the ``best'' prediction, SVM with a radial kernel, which predicts 48 out of 53 labels as false. This suggests that classification is not really working for the goal stated. To further confirm this assumption, the classification using the last 5 data points for the other measurement results is performed.

\begin{table}[]
  \centering
  \begin{tabular}{|l|l|l|l|l|l|l|}
\hline
		      & \multicolumn{2}{|l}{} & \multicolumn{1}{l}{SVM} & \multicolumn{2}{l|}{} & Naive Bayes \\ \hline
kernel		      & linear & radial & radial (tuned) & sigmoid & polynomial & - \\ \hline
cost                  & 1            & 1            & 1                        & 1             & 1                & -           \\ \hline
gamma                 & 0.01428571   & 0.01428571   & 0.1                      & 0.01428571    & 0.01428571       & -           \\ \hline
degree                & -            & -            & -                        & -             & 3                & -           \\ \hline
coef.0                & -            & -            & -                        & 0             & 0                & -           \\ \hline
\# support            & 104          & 130          & 143                      & 110           & 130              & -           \\
vectors               &              &              &                          &               &                  &             \\ \hline
accuracy              & 84.62\%      & 73.43\%      & 98.60\%                  & 58.04\%       & 76.22\%          & 66.43\%     \\
training set          &              &              &                          &               &                  &             \\ \hline
accuracy              & 52.83\%      & 62.26\%      & 62.26\%                  & 56.60\%       & 60.38\%          & 56.60\%     \\
test set              &              &              &                          &               &                  &             \\ \hline
  \end{tabular}
  \caption{Classification Results For ``Kreis19,2-2 Konzentrizitaet'' Using The Last 5 Data Points}
  \label{tab:clussclassificationresultsKreis1922Konzentrizitaetlast}
\end{table}

The classification for Kreis 19,2-2 Konzentrizitaet (Tab. \ref{tab:clussclassificationresultsKreis1922Konzentrizitaetlast}) shows that this is the same story as above. It predicts nearly everything as the label which is more common in the dataset, in this case false, to get the best result it can achieve.

\begin{table}[]
  \centering
  \begin{tabular}{|l|l|l|l|l|l|l|}
\hline
		      & \multicolumn{2}{|l}{} & \multicolumn{1}{l}{SVM} & \multicolumn{2}{l|}{} & Naive Bayes \\ \hline
kernel		      & linear & radial & radial (tuned) & sigmoid & polynomial & - \\ \hline
cost                  & 1            & 1            & 0.01                     & 1             & 1                & -           \\ \hline
gamma                 & 0.01428571   & 0.01428571   & 0.001                    & 0.01428571    & 0.01428571       & -           \\ \hline
degree                & -            & -            & -                        & -             & 3                & -           \\ \hline
coef.0                & -            & -            & -                        & 0             & 0                & -           \\ \hline
\# support            & 76           & 114          & 67                       & 70            & 111              & -           \\
vectors               &              &              &                          &               &                  &             \\ \hline
accuracy              & 88.11\%      & 81.12\%      & 78.32\%                  & 76.22\%       & 87.41\%          & 71.33\%     \\
training set          &              &              &                          &               &                  &             \\ \hline
accuracy              & 67.92\%      & 75.47\%      & 75.47\%                  & 77.36\%       & 73.58\%          & 60.38\%     \\
test set              &              &              &                          &               &                  &             \\ \hline
  \end{tabular}
  \caption{Classification Results For ``Zylinder 4,5-B Durchmesser'' Using The Last 5 Data Points}
  \label{tab:clussclassificationresultsKreisZylinder45BDurchmesserlast}
\end{table}

The classification of Zylinder 4,5-B Durchmesser (Tab. \ref{tab:clussclassificationresultsKreisZylinder45BDurchmesserlast}) works even worse than the unsuccessful classifications from above, by just predicting everything as true, although only about 77\% are labelled as such. Those results lead to the conclusion that classification for this data just does not work. One possibility could be, that the last 5 data points are just not suitable for this case. Therefore, classification is again performed for the same measurement results, but with the first 5 data points of the machining data.\\

For classification for all the measurement results using the first 5 data points the same steps as above for manual measurement 1 using the last 5 data points are carried out. The only difference is the different data set, which was used as source, which contains the first 5 data points per parameter to describe a log.

\begin{table}[]
  \centering
  \begin{tabular}{|l|l|l|l|l|l|l|}
\hline
		      & \multicolumn{2}{|l}{} & \multicolumn{1}{l}{SVM} & \multicolumn{2}{l|}{} & Naive Bayes \\ \hline
kernel		      & linear & radial & radial (tuned) & sigmoid & polynomial & - \\ \hline
cost                  & 1            & 1            & 0.01                     & 1             & 1                & -           \\ \hline
gamma                 & 0.01428571   & 0.01428571   & 0.001                    & 0.01428571    & 0.01428571       & -           \\ \hline
degree                & -            & -            & -                        & -             & 3                & -           \\ \hline
coef.0                & -            & -            & -                        & 0             & 0                & -           \\ \hline
\# support            & 40           & 84           & 33                       & 35            & 104              & -           \\
vectors               &              &              &                          &               &                  &             \\ \hline
accuracy              & 99.30\%      & 93.01\%      & 90.91\%                  & 90.21\%       & 93.71\%          & 33.57\%     \\
training set          &              &              &                          &               &                  &             \\ \hline
accuracy              & 83.02\%      & 96.23\%      & 96.23\%                  & 96.23\%       & 96.23\%          & 11.32\%     \\
test set              &              &              &                          &               &                  &             \\ \hline
  \end{tabular}
  \caption{Classification Results For ``ManualMeasurement1'' Using The First 5 Data Points}
  \label{tab:clussclassificationresultsMM1first}
\end{table}

\begin{table}[]
  \centering
  \begin{tabular}{|l|l|l|l|l|l|l|}
\hline
		      & \multicolumn{2}{|l}{} & \multicolumn{1}{l}{SVM} & \multicolumn{2}{l|}{} & Naive Bayes \\ \hline
kernel		      & linear & radial & radial (tuned) & sigmoid & polynomial & - \\ \hline
cost                  & 1            & 1            & 10                       & 1             & 1                & -           \\ \hline
gamma                 & 0.01428571   & 0.01428571   & 0.001                    & 0.01428571    & 0.01428571       & -           \\ \hline
degree                & -            & -            & -                        & -             & 3                & -           \\ \hline
coef.0                & -            & -            & -                        & 0             & 0                & -           \\ \hline
\# support            & 83           & 126          & 121                      & 117           & 136              & -           \\
vectors               &              &              &                          &               &                  &             \\ \hline
accuracy              & 88.81\%      & 88.11\%      & 82.52\%                  & 68.53\%       & 69.93\%          & 42.66\%     \\
training set          &              &              &                          &               &                  &             \\ \hline
accuracy              & 56.60\%      & 62.26\%      & 56.60\%                  & 60.38\%       & 64.15\%          & 37.74\%     \\
test set              &              &              &                          &               &                  &             \\ \hline
  \end{tabular}
  \caption{Classification Results For ``Kreis19,2-1 Konzentrizitaet'' Using The First 5 Data Points}
  \label{tab:clussclassificationresultsKreis1921Konzentrizitaetfirst}
\end{table}

\begin{table}[]
  \centering
  \begin{tabular}{|l|l|l|l|l|l|l|}
\hline
		      & \multicolumn{2}{|l}{} & \multicolumn{1}{l}{SVM} & \multicolumn{2}{l|}{} & Naive Bayes \\ \hline
kernel		      & linear & radial & radial (tuned) & sigmoid & polynomial & - \\ \hline
cost                  & 1            & 1            & 10                       & 1             & 1                & -           \\ \hline
gamma                 & 0.01428571   & 0.01428571   & 0.001                    & 0.01428571    & 0.01428571       & -           \\ \hline
degree                & -            & -            & -                        & -             & 3                & -           \\ \hline
coef.0                & -            & -            & -                        & 0             & 0                & -           \\ \hline
\# support            & 87           & 130          & 119                      & 113           & 137              & -           \\
vectors               &              &              &                          &               &                  &             \\ \hline
accuracy              & 88.81\%      & 85.31\%      & 82.52\%                  & 65.73\%       & 72.03\%          & 40.56\%     \\
training set          &              &              &                          &               &                  &             \\ \hline
accuracy              & 52.83\%      & 60.38\%      & 58.49\%                  & 60.38\%       & 62.26\%          & 39.62\%     \\
test set              &              &              &                          &               &                  &             \\ \hline
  \end{tabular}
  \caption{Classification Results For ``Kreis19,2-2 Konzentrizitaet'' Using The First 5 Data Points}
  \label{tab:clussclassificationresultsKreis1922Konzentrizitaetfirst}
\end{table}

\begin{table}[]
  \centering
  \begin{tabular}{|l|l|l|l|l|l|l|}
\hline
		      & \multicolumn{2}{|l}{} & \multicolumn{1}{l}{SVM} & \multicolumn{2}{l|}{} & Naive Bayes \\ \hline
kernel		      & linear & radial & radial (tuned) & sigmoid & polynomial & - \\ \hline
cost                  & 1            & 1            & 0.01                     & 1             & 1                & -           \\ \hline
gamma                 & 0.01428571   & 0.01428571   & 0.001                    & 0.01428571    & 0.01428571       & -           \\ \hline
degree                & -            & -            & -                        & -             & 3                & -           \\ \hline
coef.0                & -            & -            & -                        & 0             & 0                & -           \\ \hline
\# support            & 70           & 118          & 67                       & 71            & 134              & -           \\
vectors               &              &              &                          &               &                  &             \\ \hline
accuracy              & 96.50\%      & 80.42\%      & 78.32\%                  & 77.62\%       & 81.12\%          & 51.75\%     \\
training set          &              &              &                          &               &                  &             \\ \hline
accuracy              & 67.92\%      & 75.47\%      & 75.47\%                  & 75.47\%       & 75.47\%          & 49.06\%     \\
test set              &              &              &                          &               &                  &             \\ \hline
  \end{tabular}
  \caption{Classification Results For ``Zylinder 4,5-B Durchmesser'' Using The First 5 Data Points}
  \label{tab:clussclassificationresultsKreisZylinder45BDurchmesserfirst}
\end{table}

Again the results given in Tab. \ref{tab:clussclassificationresultsMM1first}, \ref{tab:clussclassificationresultsKreis1921Konzentrizitaetfirst}, \ref{tab:clussclassificationresultsKreis1922Konzentrizitaetfirst}, and, \ref{tab:clussclassificationresultsKreisZylinder45BDurchmesserfirst} indicate that, just like using the last 5 data points, all the classifications just predict everything as true or false, depending which is the majority of the data set. It seems it doesn't even matter if the first or last 5 data points are used, which strongly suggests that there is just no correlation between the machining variables and the measurement result. Just to be sure, the classification for a relatively evenly distributed measurement result is repeated using the middle 5 data points of the machining data.\\

For classification for the measurement results of Kreis 19,2-1 Konzentrizitaet using the middle 5 data points the same steps as above for manual measurement 1 using the last 5 data points are executed. The only difference is the different data set, which is used as source, which contains the middle 5 data points per parameter to describe a log.

\begin{table}[]
  \centering
  \begin{tabular}{|l|l|l|l|l|l|l|}
\hline
		      & \multicolumn{2}{|l}{} & \multicolumn{1}{l}{SVM} & \multicolumn{2}{l|}{} & Naive Bayes \\ \hline
kernel		      & linear & radial & radial (tuned) & sigmoid & polynomial & - \\ \hline
cost                  & 1            & 1            & 1                        & 1             & 1                & -           \\ \hline
gamma                 & 0.01428571   & 0.01428571   & 0.01                     & 0.01428571    & 0.01428571       & -           \\ \hline
degree                & -            & -            & -                        & -             & 3                & -           \\ \hline
coef.0                & -            & -            & -                        & 0             & 0                & -           \\ \hline
\# support            & 92           & 135          & 130                      & 118           & 136              & -           \\
vectors               &              &              &                          &               &                  &             \\ \hline
accuracy              & 88.11\%      & 76.92\%      & 71.33\%                  & 60.14\%       & 72.73\%          & 63.64\%     \\
training set          &              &              &                          &               &                  &             \\ \hline
accuracy              & 49.06\%      & 66.04\%      & 64.15\%                  & 64.15\%       & 58.49\%          & 52.83\%     \\
test set              &              &              &                          &               &                  &             \\ \hline
  \end{tabular}
  \caption{Classification Results For ``Kreis19,2-1 Konzentrizitaet'' Using The Middle 5 Data Points}
  \label{tab:clussclassificationresultsKreis1921Konzentrizitaetmiddle}
\end{table}

This classification (for results see Tab. \ref{tab:clussclassificationresultsKreis1921Konzentrizitaetmiddle}) does again not seem to work and again predicts everything as false. This again confirms the assumption that classification is not really working out for the stated goal. To eliminate another potential problem, the number of feature variables is reduced, which potentially skew the classification. Here the method described in section \ref{sec:clussfeatureselection} is used for the data sets using the last and first 5 data points for log representation.

\begin{figure}[h]
  \includegraphics[width=\textwidth]{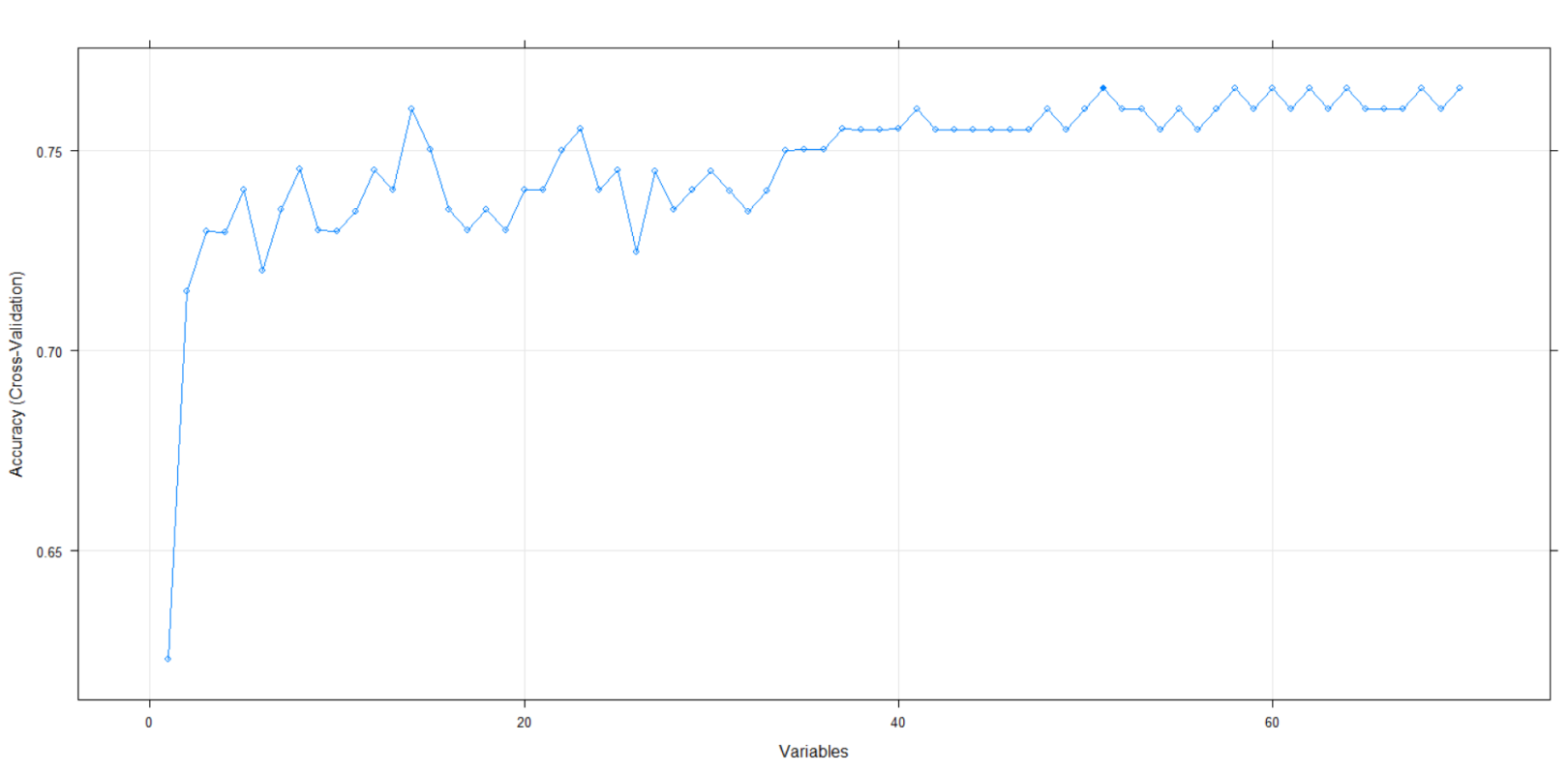}
  \caption{Recursive Feature Elimination ``Zylinder 4,5-B Durchmesser'' Last 5 Data Points}
  \label{fig:clussrecursivefeatureliminationKreisZylinder45BDurchmesserlast}
\end{figure}

The plot given in Fig. \ref{fig:clussrecursivefeatureliminationKreisZylinder45BDurchmesserlast} suggests that more feature variables don’t really lead to more accuracy at a certain point when using the last 5 data points to represent a log. Low amounts such as 14 feature variables already reach the same accuracy. Therefore, the 14 ``best'' features are used to create a new data frame and perform classification. The 14 feature variables are (the number after the dot indicates which of the values within the 5 values chosen - this means that ``.1'' is the 5th value and ``.5'' is the first value counting from the last of the five values):
\begin{itemize}
  \item ns=2;s=/Channel/MachineAxis/aaLoad[u1,2].1
  \item ns=2;s=/Channel/MachineAxis/aaLoad[u1,2].2
  \item ns=2;s=/Channel/MachineAxis/aaLeadP[u1,1].5
  \item ns=2;s=/Channel/MachineAxis/aaLeadP[u1,1].4
  \item ns=2;s=/Channel/MachineAxis/aaLoad[u1,2].4
  \item ns=2;s=/Channel/MachineAxis/aaLoad[u1,2].3
  \item ns=2;s=/Channel/Spindle/driveLoad.3
  \item ns=2;s=/Channel/MachineAxis/aaTorque[u1,3].5
  \item ns=2;s=/Channel/MachineAxis/aaTorque[u1,3].3
  \item ns=2;s=/Channel/MachineAxis/aaTorque[u1,2].3
  \item ns=2;s=/Channel/MachineAxis/aaLoad[u1,3].3
  \item ns=2;s=/Channel/MachineAxis/aaLoad[u1,3].1
  \item ns=2;s=/Channel/MachineAxis/aaTorque[u1,2].5
  \item ns=2;s=/Channel/MachineAxis/aaLoad[u1,1].4
\end{itemize}

\begin{table}[]
  \centering
  \begin{tabular}{|l|l|l|l|l|l|l|}
\hline
		      & \multicolumn{2}{|l}{} & \multicolumn{1}{l}{SVM} & \multicolumn{2}{l|}{} & Naive Bayes \\ \hline
kernel		      & linear & radial & radial (tuned) & sigmoid & polynomial & - \\ \hline
cost                  & 1            & 1            & 1                        & 1             & 1                & -           \\ \hline
gamma                 & 0.07142857   & 0.07142857   & 1                        & 0.07142857    & 0.07142857       & -           \\ \hline
degree                & -            & -            & -                        & -             & 3                & -           \\ \hline
coef.0                & -            & -            & -                        & 0             & 0                & -           \\ \hline
\# support            & 80           & 87           & 98                       & 63            & 78               & -           \\
vectors               &              &              &                          &               &                  &             \\ \hline
accuracy              & 78.32\%      & 79.02\%      & 87.41\%                  & 75.52\%       & 82.52\%          & 69.23\%     \\
training set          &              &              &                          &               &                  &             \\ \hline
accuracy              & 75.47\%      & 75.47\%      & 75.47\%                  & 75.47\%       & 73.58\%          & 60.38\%     \\
test set              &              &              &                          &               &                  &             \\ \hline
  \end{tabular}
  \caption{Classification Results For ``Zylinder 4,5-B Durchmesser'' With Feature Elimination Using The Last 5 Data Points}
  \label{tab:clussclassificationresultsKreisZylinder45BDurchmesserlastfeatureelimination}
\end{table}

Classification results given in Tab. \ref{tab:clussclassificationresultsKreisZylinder45BDurchmesserlastfeatureelimination} for ``Zylinder 4,5-B Durchmesser'' with feature elimination using the last 5 data points shows that even having less data points, nothing changed. Again the prediction is really one sided. To confirm that feature elimination doesn't change the success of classification, feature elimination is also performed for the first 5 data points.

\begin{figure}[h]
  \includegraphics[width=\textwidth]{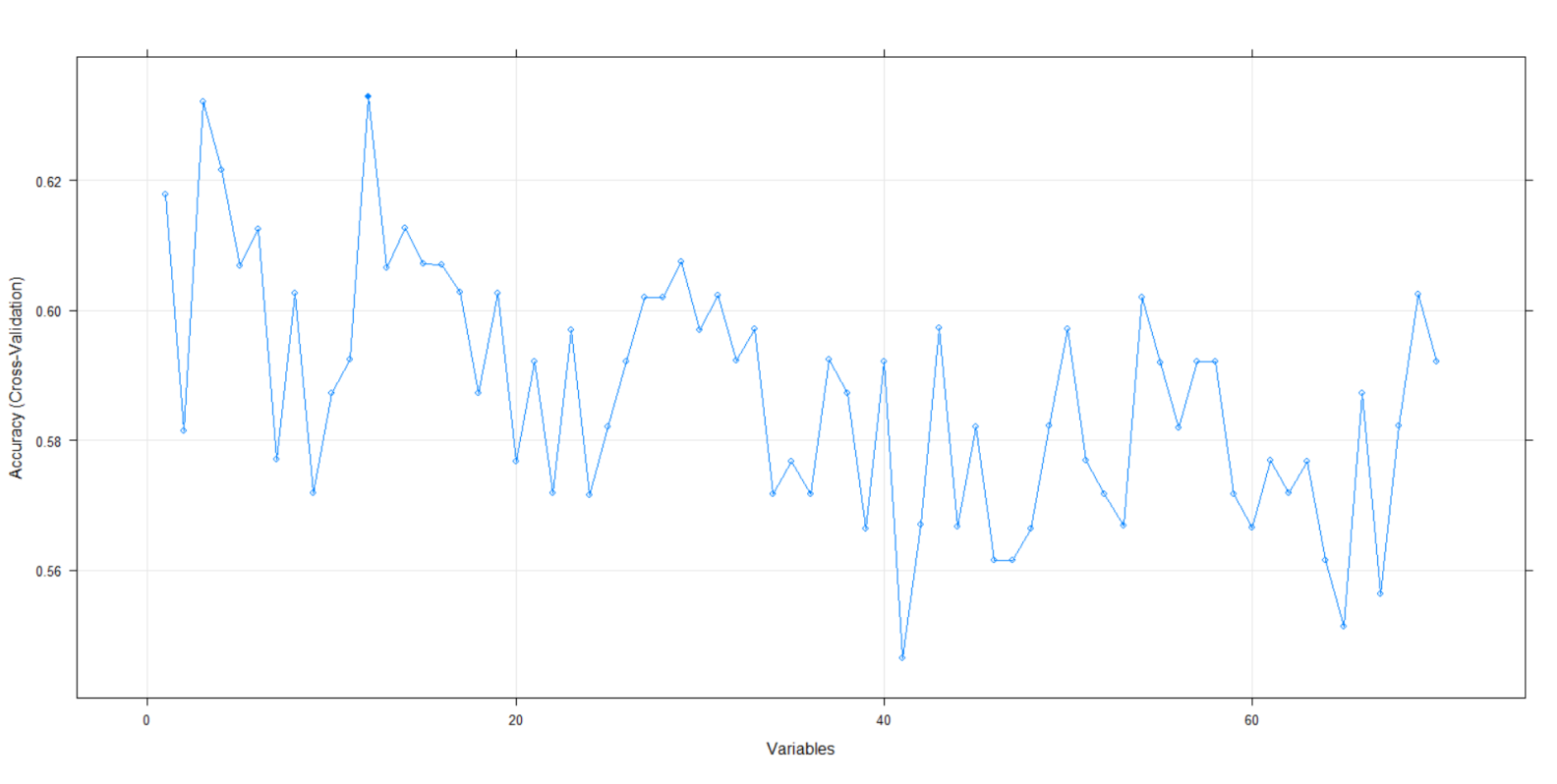}
  \caption{Recursive Feature Elimination ``Kreis19,2-2 Konzentrizitaet'' First 5 Data Points}
  \label{fig:clussrecursivefeatureliminationKreisZylinder45BDurchmesserfirst}
\end{figure}

The plot given in Fig. \ref{fig:clussrecursivefeatureliminationKreisZylinder45BDurchmesserfirst} suggests that less feature variables lead to more accuracy up to a certain point when using the first 5 data points to represent a log. Low amounts such as 12 feature variables already reach the best accuracy. These 12 feature variables are (the number after the dot indicates which of the values within the 5 values chosen -this means that ``.1'' is the 5th value and ``.5'' is the first value counting from the last of the five values):
\begin{itemize}
  \item ns=2;s=/Channel/MachineAxis/aaTorque[u1,1].5
  \item ns=2;s=/Channel/MachineAxis/aaLoad[u1,2].3
  \item ns=2;s=/Channel/MachineAxis/aaTorque[u1,3].5
  \item ns=2;s=/Channel/MachineAxis/aaLoad[u1,1].2
  \item ns=2;s=/Channel/MachineAxis/aaTorque[u1,2].1
  \item ns=2;s=/Channel/MachineAxis/aaTorque[u1,1].4
  \item ns=2;s=/Channel/MachineAxis/aaLeadP[u1,1].4
  \item ns=2;s=/Channel/MachineAxis/aaLoad[u1,1].3
  \item ns=2;s=/Channel/MachineAxis/aaLoad[u1,2].5
  \item ns=2;s=/Channel/MachineAxis/aaTorque[u1,1].3
  \item ns=2;s=/Channel/MachineAxis/aaVactB[u1,3].5
  \item ns=2;s=/Channel/MachineAxis/aaTorque[u1,2].5
\end{itemize}

\begin{table}[]
  \centering
  \begin{tabular}{|l|l|l|l|l|l|l|}
\hline
		      & \multicolumn{2}{|l}{} & \multicolumn{1}{l}{SVM} & \multicolumn{2}{l|}{} & Naive Bayes \\ \hline
kernel		      & linear & radial & radial (tuned) & sigmoid & polynomial & - \\ \hline
cost                  & 1            & 1            & 1                        & 1             & 1                & -           \\ \hline
gamma                 & 0.08333333   & 0.08333333   & 0.01                     & 0.08333333    & 0.08333333       & -           \\ \hline
degree                & -            & -            & -                        & -             & 3                & -           \\ \hline
coef.0                & -            & -            & -                        & 0             & 0                & -           \\ \hline
\# support            & 103          & 125          & 112                      & 110           & 113              & -           \\
vectors               &              &              &                          &               &                  &             \\ \hline
accuracy              & 69.23\%      & 76.92\%      & 64.34\%                  & 58.74\%       & 73.43\%          & 67.13\%     \\
training set          &              &              &                          &               &                  &             \\ \hline
accuracy              & 62.26\%      & 56.60\%      & 62.26\%                  & 62.26\%       & 60.38\%          & 67.92\%     \\
test set              &              &              &                          &               &                  &             \\ \hline
  \end{tabular}
  \caption{Classification Results For ``Kreis19,2-2 Konzentrizitaet'' With Feature Elimination Using The First 5 Data Points}
  \label{tab:clussclassificationresultsKreisZylinder45BDurchmesserfirstfeatureelimination}
\end{table}

Removing feature variables in this approach leads to an interesting effect shown in Tab. \ref{tab:clussclassificationresultsKreisZylinder45BDurchmesserfirstfeatureelimination}. Instead of just going the safe route of predicting everything false, it predicts more wrongfully true, which naturally leads to a lower accuracy, than using all feature variables. This is the last attempt to predict the outcome of the measurement using only the machining data via classification. There could be certain data points which would work, but for this relatively generic way, the solution is not useful.\\

Since the first goal isn’t achieved, a new approach is chosen, that looks if the machining data can predict the outcome of not only a single measurement, but the whole part. A part where the measurement of at least one automatic measurement, except the first three (the first three being the ones starting with ``Flaeche'' in Fig. \ref{fig:teststats}), is NOK/FALSE is considered faulty. Again the same classification method as above is used. The classification is performed using the last 5, the first 5 and the middle 5 data points with naive bayes and different kernels for SVM.

\begin{table}[]
  \centering
  \begin{tabular}{|l|l|l|l|l|l|l|}
\hline
		      & \multicolumn{2}{|l}{} & \multicolumn{1}{l}{SVM} & \multicolumn{2}{l|}{} & Naive Bayes \\ \hline
kernel		      & linear & radial & radial (tuned) & sigmoid & polynomial & - \\ \hline
cost                  & 1            & 1            & 0.01                     & 1             & 1                & -           \\ \hline
gamma                 & 0.01428571   & 0.01428571   & 0.001                    & 0.01428571    & 0.01428571       & -           \\ \hline
degree                & -            & -            & -                        & -             & 3                & -           \\ \hline
coef.0                & -            & -            & -                        & 0             & 0                & -           \\ \hline
\# support            & 89           & 125          & 88                       & 91            & 121              & -           \\
vectors               &              &              &                          &               &                  &             \\ \hline
accuracy              & 86.01\%      & 77.62\%      & 72.03\%                  & 70.63\%       & 81.82\%          & 70.63\%     \\
training set          &              &              &                          &               &                  &             \\ \hline
accuracy              & 64.15\%      & 71.70\%      & 71.70\%                  & 69.81\%       & 67.92\%          & 66.04\%     \\
test set              &              &              &                          &               &                  &             \\ \hline
  \end{tabular}
  \caption{Classification Results For Overall Automatic Measurement Using The Last 5 Data Points}
  \label{tab:clussclassificationresultsautomaticmeasurementlast}
\end{table}

\begin{table}[]
  \centering
  \begin{tabular}{|l|l|l|l|l|l|l|}
\hline
		      & \multicolumn{2}{|l}{} & \multicolumn{1}{l}{SVM} & \multicolumn{2}{l|}{} & Naive Bayes \\ \hline
kernel		      & linear & radial & radial (tuned) & sigmoid & polynomial & - \\ \hline
cost                  & 1            & 1            & 0.01                     & 1             & 1                & -           \\ \hline
gamma                 & 0.01428571   & 0.01428571   & 0.001                    & 0.01428571    & 0.01428571       & -           \\ \hline
degree                & -            & -            & -                        & -             & 3                & -           \\ \hline
coef.0                & -            & -            & -                        & 0             & 0                & -           \\ \hline
\# support            & 78           & 121          & 86                       & 86            & 133              & -           \\
vectors               &              &              &                          &               &                  &             \\ \hline
accuracy              & 90.91\%      & 81.12\%      & 72.03\%                  & 72.03\%       & 78.32\%          & 36.36\%     \\
training set          &              &              &                          &               &                  &             \\ \hline
accuracy              & 67.93\%      & 71.70\%      & 71.70\%                  & 69.81\%       & 71.70\%          & 32.08\%     \\
test set              &              &              &                          &               &                  &             \\ \hline
  \end{tabular}
  \caption{Classification Results For Overall Automatic Measurement Using The First 5 Data Points}
  \label{tab:clussclassificationresultsautomaticmeasurementfirst}
\end{table}

\begin{table}[]
  \centering
  \begin{tabular}{|l|l|l|l|l|l|l|}
\hline
		      & \multicolumn{2}{|l}{} & \multicolumn{1}{l}{SVM} & \multicolumn{2}{l|}{} & Naive Bayes \\ \hline
kernel		      & linear & radial & radial (tuned) & sigmoid & polynomial & - \\ \hline
cost                  & 1            & 1            & 0.01                     & 1             & 1                & -           \\ \hline
gamma                 & 0.01428571   & 0.01428571   & 0.001                    & 0.01428571    & 0.01428571       & -           \\ \hline
degree                & -            & -            & -                        & -             & 3                & -           \\ \hline
coef.0                & -            & -            & -                        & 0             & 0                & -           \\ \hline
\# support            & 83           & 131          & 89                       & 91            & 121              & -           \\
vectors               &              &              &                          &               &                  &             \\ \hline
accuracy              & 88.81\%      & 77.62\%      & 72.03\%                  & 72.03\%       & 80.42\%          & 64.34\%     \\
training set          &              &              &                          &               &                  &             \\ \hline
accuracy              & 60.38\%      & 71.70\%      & 71.70\%                  & 75.47\%       & 67.92\%          & 50.94\%     \\
test set              &              &              &                          &               &                  &             \\ \hline
  \end{tabular}
  \caption{Classification Results For Overall Automatic Measurement Using The Middle 5 Data Points}
  \label{tab:clussclassificationresultsautomaticmeasurementmiddle}
\end{table}

Just like the classification for single measurement results, the SVM predicts everything as the majority of the labels (the test set is 75\% false), which is just not a useable way for classification. Furthermore, no matter which data points are used, the results given in Tab. \ref{tab:clussclassificationresultsautomaticmeasurementlast}, \ref{tab:clussclassificationresultsautomaticmeasurementfirst}, and \ref{tab:clussclassificationresultsautomaticmeasurementmiddle} are very similar, which speaks for no correlation between the data points and the actual outcome of the part.\\

Another approach used is to look at manual measurements and automatic measurements and compare the results (as described above for the manual measurement to be OK/TRUE all three manual measurements have to be OK/TRUE and for the automatic measurement all but the first three need to be OK/TRUE).

\begin{table}[]
  \centering
  \begin{tabular}{|l|l|l|}
\hline
                 & MM\_true & MM\_false \\ \hline
automatic\_true  & 51       & 4         \\ \hline
automatic\_false & 129      & 12        \\ \hline
  \end{tabular}
  \caption{Comparison Of Manual And Automatic Measurement}
  \label{tab:clusscomparisonmanualandautomaticmeasurement}
\end{table}

As can be seen in Tab. \ref{tab:clusscomparisonmanualandautomaticmeasurement}, 12 of the 16 manual measurement results being NOK are actually NOK (with 4 being actually OK) and 51 of the 180 manual results being OK are actually OK (with 129 being actually NOK). This applies only under the assumption that the automatic measurement is always correct. These results are not good enough to predict the outcome of the actual (automatic measurement) result but it can definitely be used for some considerations. Thinking of a scenario where the automatic measurement is very expensive and the production of one part is quite cheap it might for example be a good idea to just make an automatic quality control for parts that pass the manual one and accept that with this approach some good parts might not be sold but nonetheless the overall revenue might be higher. More economic considerations with the goal to optimise the cash flow can be performed with further knowledge of the actual prices for measurements and production.

\subsection{Lessons Learned}
Overall, classification and clustering of the data achieves good results for the first scenario looking at it from the point of accuracy. A major problem with these results is that they seem unstable (e.g. looking at the different results for different seed values regarding the split between training and test data set) which is most likely based on the small number of logs available. To make a final conclusion on the precision of clustering and classification approaches using this data a higher number of logs would be needed. Additionally, there is the problem of not all logs measuring the same variables. If this was more standardized, it would be possible to use a higher number of variables for analysis which would probably make the results even better. All in all, the results obtained meet the goal of describing the data given pretty well but the question is if this result is meaningful as this would require a higher number of logs.

For, the second scenario classification does not provide satisfactory results looking at the goal of using the machining data to predict measurement results or the overall “OKness” of a part. There are quite a few reasons that could be made responsible for this lack of accuracy for the classification methods all having as root cause that the wrong machining data is used for prediction with the most obvious one being that 5 data points per parameter do simply not contain enough information to make a meaningful prediction. Another reason could be that due to trying it with only 3 different quintetts (first 5, middle 5, last 5) the region that is important for the quality of the part is not covered with the data selected. Furthermore, it is important to also take into consideration that it might not be possible to predict product quality based on the machining data provided at all.

A further difficulty in performing classification is the ratio of passed to not passed tests for most of the measurements - as discussed in the beginning of section \ref{sec:clussscenario2}, having a very high percentage of parts failing or succeeding at a test makes it impossible to obtain a feasible prediction formula as just predicting the result which occurs in nearly all cases provides a very high accuracy.

Another interesting point which is discussed in greater detail in the last paragraphs of section \ref{sec:clussscenario2} is the role manual measuring has when it comes to determining the quality of parts. Even if manual measuring does not really give a hint about how the result of the automatic measurement will be, it could be useful for taking economic decisions. It might for example be feasible to decide if the automatic measurements should be carried out based on the manual measurement. The outcome of this consideration depends strongly on the cost of manufacturing per part, the cost of automatic measurements as well as of the costs that are anticipated for selling faulty parts.


\section{Conclusion and Outlook} 
\label{sec:conclusion}

This work shows how conformance checking, classification, and clustering of machining data gathered by executing a BPMN manufacturing process with a workflow engine can be performed. Furthermore, the results achieved are shown and discussed with respect to their quality, possible improvements, and suggestions for future work.
Overall, the results show that data collected within a manufacturing process can be used for process-oriented analysis such as conformance checking of process logs or classification and clustering of the processes based on machining data in order to predict if produced parts are good or bad. In contrast to the resource-based data collection method where data-streams of single machines have to be saved in databases, cleaned and re-contextualized by connecting it to orders, batches, or single products, the aforementioned method has the advantage of providing data for the whole BPMN process which makes it easy to retrieve data for individual parts for further analysis.
Future work in this field will include visualizing the data contained within the process logs, finding standardized ways to represent logs of individual parts to make it possible to compare the machining data points of different parts, and determining ways to choose parameters that should be recorded during the process execution. Additionally, it would be desirable to evaluate whether the ex-post analysis performed in this paper can also be undertaken at run-time.


~\\
\textbf{Acknowledgements:} This work has been partially supported and funded by
the Austrian Research Promotion Agency (FFG) via the “Austrian Competence
Center for Digital Production” (CDP) under the contract number 854187.

\bibliographystyle{splncs03}
\bibliography{refs.bib}

\end{document}